# Temperature-gradient induced massive augmentation of solute dispersion in viscoelastic micro-flows

Siddhartha Mukherjee,[1], Sunando DasGupta[1,2], Suman Chakraborty[1,3*]

[1]*Advanced Technology Development Center, Indian Institute of Technology Kharagpur, Kharagpur, India-721302*

[2]*Department of Chemical Engineering, Indian Institute of Technology Kharagpur, Kharagpur, India-721302*

[3]*Department of Mechanical Engineering, Indian Institute of Technology Kharagpur, Kharagpur, India-721302*

Enhancing solute dispersion in electrically actuated flows has always been a challenging proposition, as attributed to the inherent uniformity of the flow field in absence of surface patterns. Over the years, researchers have focused their attention towards circumventing this limitation, by employing several fluidic and geometric modulations. However, the corresponding improvements in solute dispersion often turn out to be inconsequential. Here we unveil that by exploiting the interplay between an externally imposed temperature gradient, subsequent electrical charge redistribution and ionic motion, coupled with the rheological complexities of the fluid, one can achieve up to one order of magnitude enhancement of solute dispersion in a pressure-driven flow of an electrolyte solution. Our results demonstrate that the complex coupling between thermal, electrical, hydro-dynamic and rheological parameters over small scales, responsible for such exclusive phenomenon, can be utilitarian in designing novel thermally-actuated micro and bio-microfluidic devices with favorable solute separation and dispersion characteristics.

---

[*]Corresponding author, email: suman@mech.iitkgp.ernet.in

## 1. Introduction

Integrating multiple fluidic processes into a single platform has become progressively important in modern lab-on-a-chip devices where separation and mixing often turn out to be two most critical processes (Ghosal 2004; Hunter 1981; J H Masliyah & S Bhattacherjee 2006; R. F. Probstein 1994; Stone *et al.* 2004; Stroock *et al.* 2002; Whitesides 2006). With rapid advancement in micro-fabrication technologies, a large number of research efforts have been dedicated towards developing strategies for improved fluidic mixing or separation (Anderson *et al.* 2000; Chang & Yang 2008; Ghosh & Chakraborty 2012; Glasgow *et al.* 2004; Karniadakis G. 2005; Sugioka 2010; Zhang *et al.* 2006). Towards achieving enhanced mixing in micro-devices, diffusion and dispersion are undoubtedly the two most common phenomena. Accordingly, significant research interest in this domain has evolved over the past years, with a vision of employing different flow actuating mechanisms as well as geometric alterations in the fluidic pathways, so as to achieve the desired functionalities.

Although flow actuation using electric fields has wide spectrum of applications in both engineering and medical domain (Bandopadhyay & Chakraborty 2012; Becker & Gärtner 2000; Berli 2010; Das *et al.* 2006, 2018; Das S., and Chakraborty S. 2007; Garcia *et al.* 2005; Haeberle & Zengerle 2007; Mandal *et al.* 2012; Mark *et al.* 2010; Nguyen *et al.* 2013; Ohno *et al.* 2008; Van Der Heyden *et al.* 2005; Zhao 2011), one of the major aspects of the classical electroosmotic flow (EOF) in presence of homogenous interfacial conditions is the existence of the uniform velocity profile which arises when the electrical double layer (EDL) becomes very thin compared to the channel dimension (Ghosal 2004; J H Masliyah & S Bhattacherjee 2006). This results in a plug-type velocity distribution, thus reducing the extent of mixing significantly.

Hydrodynamic dispersion is the band broadening of the solute which mainly arises from the non-uniformity in the flow field (Ajdari *et al.* 2006; Arcos *et al.* 2018; Aris 1956, 1959; Barton 1983; Chatwin 1970, 1975; Chatwin & Sullivan 1982; Datta & Ghosal 2008; Dutta 2008; Ghosal 2006; Ghosal & Chen 2012; Jansons 2006; Mazumder & Das 1992; Ng & Yip 2001; Rana & Murthy 2016; Smith 1982; Sounart & Baygents 2007; Taylor 1953; Watson 1983; Zholkovskij & Masliyah 2004). Under ideal circumstances, velocity profile of EOF does not contribute to the shear-induced axial dispersion because of the flatness of the velocity profile as opposed to the case of Poiseuille flow (which is parabolic in nature) (Gaš *et al.* 1997; Ghosal 2004; Mukherjee *et al.* 2019). However, in practice, any in-homogeneity in the flow condition or flow domain can give rise to strong perturbation in the flow field, thereby inducing an axial pressure-gradient, which is accompanied by the generation of secondary flow component in order to maintain the flow continuity. In applications demanding augmented dispersion, classical EOF is modulated in two ways, either bringing non-uniformity in the channel geometry or introducing axial variation in the zeta-potential (Ajdari 1995, 1996; Arcos *et al.* 2018; Ghosal 2002; Ghosh *et al.* 2017; Mandal *et al.* 2015).

Over the years, conventional studies on electrokinetics mainly directed their focus towards different techniques of flow actuation, energy conversion and zeta potential measurement under isothermal flow condition (Bandopadhyay & Chakraborty 2011; Brask *et al.* 2005; Choi *et al.* 2011; Das & Chakraborty 2010; Gao *et al.* 2005; Levine *et al.* 1975; Li *et al.* 2009, 2011; Mogensen *et al.* 2009; Sinton *et al.* 2002; Tandon *et al.* 2008; Venditti *et al.* 2006; Zeng *et al.* 2001). The corresponding literature for non-isothermal flow is relatively scarce because of the lack of understanding of the physics involved. In non-isothermal systems, several complexities come into picture. First, the modulated thermo-physical properties of electrolyte solution like viscosity, electrical permittivity, thermal conductivity, ionic diffusivity, thermophoretic mobility, in presence of a thermal gradient, strongly influence the fluid motion. Besides this alteration in hydrodynamics, an additional contribution of dielectrophoretic body force due to permittivity variation, accompanied by an induced axial pressure gradient, comes into existence in addition to the conventional electrokinetic forcing, thereby bringing complexity in the flow physics (Dietzel & Hardt 2017; Ghonge *et al.* 2013). Additionally, zeta-potential, which plays a crucial dole in governing the flow physics in electrokinetic flows, no longer remains constant in presence of a thermal gradient (Ghonge *et al.* 2013; Reppert 2003; Revil *et al.* 1999, 2003; Venditti *et al.* 2006). Also, the determination of the temperature field may require the knowledge of other effects like heat generation due to induced streaming field or viscous dissipation which in turn affect fluid physical properties and convective contribution to the temperature distribution. Apart from this, contrary to the conventional electrokinetic studies, the assumption of mechanical equilibrium of ions within the electrical double layer (EDL) (under isothermal condition) is no longer valid where the effect of thermo-diffusion of ions needs to be incorporated in the transport equation of ionic species along with other components (Zhang *et al.* 2019; Zhou *et al.* 2015).

The distortion of the local equilibrium of ions in the EDL creates a departure from the classical Boltzmann distribution of ions which has direct consequences on the potential distribution. This alteration in charge distribution influences the flow dynamics via electrokinetic forcing, where the intricate coupling between thermal and electrical effects are already prevalent through the aforesaid property variation and an additional dielectrophoretic force. Further, in the absence of any external electric field, the net ionic current turns out to be zero. This condition is itself another source of non-linearity in the analysis where both conduction and streaming current undergo drastic alteration under the influence of finite temperature difference. In addition, it is noteworthy to mention that, when an electrolyte solution is subjected to an imposed temperature gradient, a thermo-electric field is induced by virtue of the movement of ions in response to the thermal driving force, commonly known as Soret effect (Dietzel & Hardt 2016, 2017; Ghonge *et al.* 2013; Maheedhara *et al.* 2018; Zhang *et al.* 2019; Zhou *et al.* 2015). Moreover, additional form of thermo-electric field can be induced within the system because of the diffusivity difference between the ions even if their Soret coefficients remain the same. Besides, the mode of application of thermal gradient can cause significant rearrangement of ions within the EDL and hence, the subsequent potential distribution for a transverse temperature gradient may not necessarily be same with that of longitudinal temperature, altering the hydrodynamics in a rather profound manner. Considering the aforementioned intricacies in coupling thermal, electrical and hydro-dynamical effects in micro-confinements, research efforts towards addressing various aspects of thermo-solutal convection of electrolyte solutions have turned out to be relatively inadequate, despite having widespread applications in processes like water treatment, charge separation, zeta-potential determination, waste heat recovery, and energy conversion (Barragán & Kjelstrup 2017; Dietzel & Hardt 2016, 2017; Jokinen *et al.* 2016; Li & Wang 2018; Sandbakk *et al.* 2013; Würger 2008, 2010; Xie *et al.* 2018). As such, the research focus in this domain has been directed primarily towards incorporating non-isothermal effects as a secondary force in the alteration of hydrodynamics of simple fluids (Chakraborty 2006; Garai & Chakraborty 2009; Huang & Yang 2006; Keramati *et al.* 2016; Maynes & Webb 2003; Sadeghi *et al.* 2011; Sánchez *et al.* 2018; Tang *et al.* 2003; Xuan *et al.* 2004a; Xuan 2008; Xuan *et al.* 2004b; Yavari *et al.* 2012). Therefore, except for some limited physical scenarios, however, such an exclusive effect has not been utilized to a significant practical benefit (Dietzel & Hardt 2016, 2017; Zhang *et al.* 2019).

Recently, incorporation of thermal gradient has emerged as an alternative tool in augmenting dispersion where interplay between thermal and electrical effects over small length scales, almost exclusively, dictates the flow physics (Chen *et al.* 2005; Mukherjee *et al.* 2019; Sánchez *et al.* 2018). In addition, it may be noted that the emergence of new generation medical devices, complex bio-fluids have more prominently come into the paradigm of microfluidics (Berli 2010; Berli & Olivares 2008; Das & Chakraborty 2006; Olivares *et al.* 2009; Zhao & Yang 2011, 2013). Such fluids exhibit strikingly distinct behavior compared to the fluids obeying Newton's law of viscosity (Brust *et al.* 2013; De Loubens *et al.* 2011; Fam *et al.* 2007; Moyers-Gonzalez *et al.* 2008; Owens 2006; Silva *et al.* 2017; Vissink *et al.* 1984). Some recent studies have demonstrated that the constitutive behavior of these biological fluids has close resemblance with the rheology of viscoelastic fluids and therefore, the inclusion of fluid rheology and viscoelasticity in dispersion characteristics has gained significant attention lately (Arcos *et al.* 2018; Brust *et al.* 2013; Hoshyargar *et al.* 2018; Mukherjee *et al.* 2019). While considering the thermally induced electrokinetic flow of viscoelastic fluids, an additional source of non-linearity crops in as mediated by the constitutive behavior of the fluid (Afonso *et al.* 2009, 2013; Coelho *et al.* 2012; Ferrás *et al.* 2016; Ghosh *et al.* 2016; Ghosh & Chakraborty 2015). Moreover, the degree of viscoelasticity, which is determined by using physical properties like fluid viscosity and relaxation time, is a strong function of the prevalent thermal gradient, augmenting the complexity of the problem to a large extent (Bautista *et al.* 2013; Mukherjee *et al.* 2019).

To the best of our knowledge, dispersion characteristics of thermally induced electrokinetic transport of complex fluids in microfluidic environment, where the temperature gradient is solely used for flow manipulation, has not been addressed in the literature. Here, we report the effect of an external temperature gradient on the dispersion characteristics of an electrolyte solution in a parallel plate microchannel. We subsequently discuss the charge redistribution upon applied thermal gradient, subsequent perturbation on the fluid motion and its implications on hydrodynamic dispersion, considering both Newtonian and viscoelastic fluids. Our results reveal that by combining some of the electrokinetic, thermal and fluidic parameters coupled with rheological aspects, it is possible to achieve massively augmented solute dispersion, while for some combinations significant enhancement in streaming potential (compared to the solely pressure-driven flow) can also be obtained. We believe that the present analysis can be used as a fundamental basis in the design of thermally actuated micro and bio-fluidic devices demanding improved solute dispersion where the interplay between electro-mechanics, thermal effect, hydrodynamic and rheological aspects in narrow confinement can be coupled together to a beneficial effect.

## 2. Problem formulation

We consider non-isothermal electrokinetic flow of a binary 1:1 symmetric electrolyte solution through a parallel plate microchannel. We choose rectangular Cartesian co-ordinate system where $x$ and $y$ co-ordinates represent longitudinal and transverse directions respectively while the origin is placed at the centreline of the channel. Length scales in the two directions are $l$ and $h$ respectively where the half-channel height ($h$) is very small compared to the channel length ($l$), i.e. $h << l$ or $\beta = h / l << 1$. We have employed two different types of thermal gradients, (a) Case 1: axially applied temperature gradient and (b) Case 2: temperature gradient applied in the transverse direction. First, we will briefly discuss about the formulation of the main focus of the present analysis, i.e. obtaining the hydrodynamic dispersion coefficient which directly depends on the flow velocity which in turn is influenced strongly by the charge distribution modulated by external temperature gradient. Hence, formulation regarding temperature field is presented first and the resulting potential distribution and velocity fields are presented subsequently. Final part of the present analysis is the inclusion of fluid rheology to examine its role in altering the dispersion characteristics.

### 2.1 *Dispersion coefficient*

We consider the hydrodynamic dispersion characteristics of a combined pressure-driven and thermal-gradient driven electrokinetic flow of an electrolyte solution. According to the definition of band-broadening phenomenon, hydrodynamic dispersion coefficient ($\bar{D}_{eff}$) depends on the rate of change of variance of solute displacement band as $D_{eff} = \frac{1}{2}\frac{d}{dt}\sigma^2(t)$ where $\sigma^2$ is the variance of solute displacement. This temporal variation with respect to the centroid of the band depends on the plate height ($h'$) as $h' = \frac{d}{dx'}\sigma^2(x')$ where $x'$ denotes the location of the band centroid. The knowledge of the plate height is necessary in determining the dispersion coefficient because of

(a)

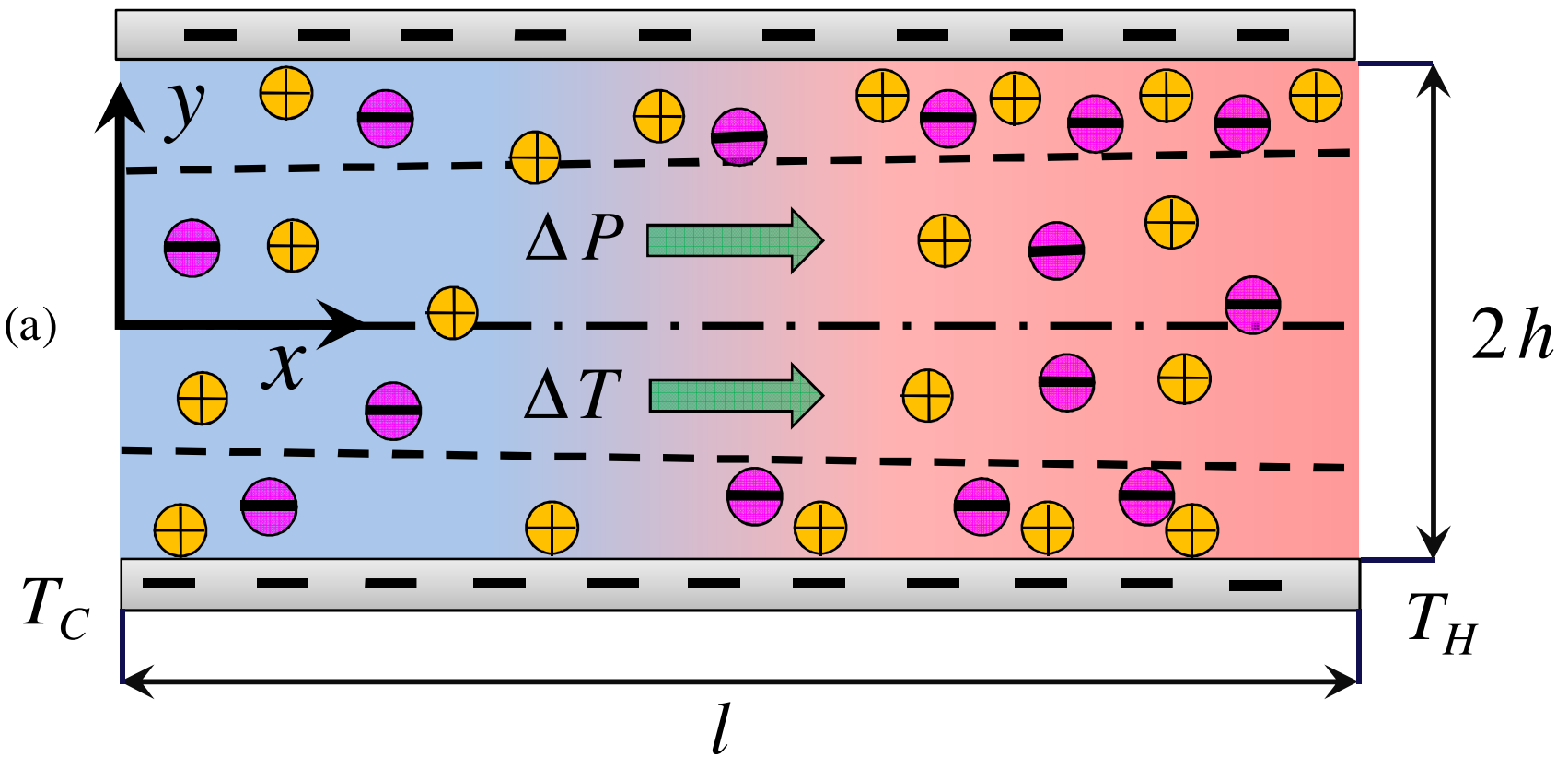

(b)

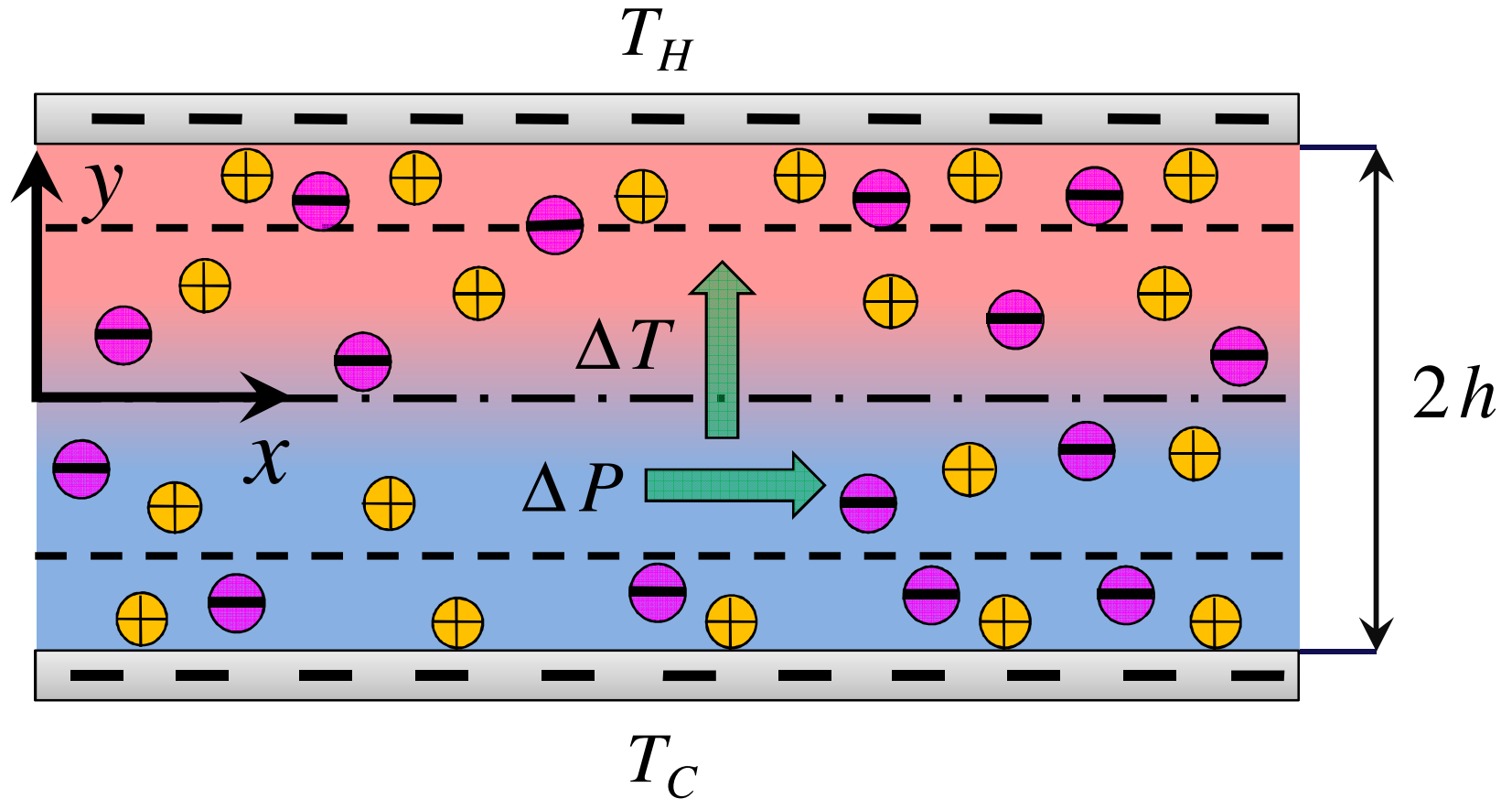

FIGURE 1. Schematic of the combined pressure-driven and temperature-gradient induced flow of electrolyte solution through a parallel plate microchannel. (a) Temperature gradient is applied in the axial direction, (b) temperature gradient is applied in the transverse direction.

its ability to incorporate any change in the variance of the mean concentration. For a band of non-adsorbing solute flowing in a rectangular channel, the velocity of the centre mass becomes equal to the mean velocity of flow, i.e. $u_{avg} = dx'/dt$. Using the descriptions of $h'$ and $x'$, one can rewrite the expression of dispersion coefficient as

$$D_{eff} = u_{avg}\, h'/2 \tag{1}$$

Here, $u_{avg}$ is the mean velocity averaged cross-sectionally. As reported in the literature, (van Deemter *et al.* 1956), the plate height is related to the mean velocity a

$$h' = 2D/u_{avg} + \left(u_{avg}\, h_{\min}{}^{2}/8D\right) \tag{2}$$

In equation (2), $h_{min}$ is the minimum plate height for a given flow condition and $D$ is diffusivity. On the right side, the contribution of the molecular diffusion is represented by the first term while the second term indicates the contribution due to the non-uniformity in the flow field. By following some recent studies (Arcos *et al.* 2018; Hoshyargar *et al.* 2018; Zholkovskij & Masliyah 2006), we have used the following expression for evaluating $h_{min}$

$$h_{\min}{}^{2}=\frac{16}{h}\int_0^h\int_0^y\left[\left\{\left(u/u_{avg}\right)-1\right\}dy\right]^2 dy \tag{3}$$

Combining all this, the dimensionless form of dispersion coefficient ($\bar{D}_{eff}$) reads as

$$\bar{D}_{eff}=1+\left(Pe_D\,\bar{u}_{avg}\,\bar{h}_{\min}\right)^2/16 \tag{4}$$

where $Pe_D$ is the Dispersion Peclet number, $\bar{h}_{\min}=h_{\min}/h$ and $\bar{u}_{avg}=u_{avg}/u_c$ is the dimensionless average flow velocity with $u_c$ being the characteristic velocity scale.

2.2 *Temperature distribution*

Unlike conventional streaming field induced electrokinetic flow, here temperature ($T$) within the microchannel does not remain constant which is given by the energy equation:

$$\rho C_p\left(u\frac{\partial T}{\partial x}+v\frac{\partial T}{\partial y}\right)=\frac{\partial}{\partial x}\left(k\frac{\partial T}{\partial x}\right)+\frac{\partial}{\partial y}\left(k\frac{\partial T}{\partial y}\right)+Q_{gen}+Q_{vd} \tag{5}$$

where $\rho$, $C_p$ and $k$ are density, specific heat capacity and thermal conductivity of the fluid respectively. The left hand side of equation (5) represents the advective component of the thermal transport, whereas the first and second terms on the right side represent the axial and transverse conductive components. $Q_{gen}$ is the heat generation term due to the induced streaming field and $Q_{vd}$ is the viscous dissipation term. $Q_{gen}$ can be expressed as $Q_{gen}\sim\sigma E_x{}^2$ where $E_x=-\partial\phi/\partial x$ is the induced streaming field and $\sigma$ bulk electrical conductivity, $\sigma=2z^2e^2Dn_\infty/k_BT$ with $z$, $e$, $D$, $n_\infty$ and $k_B$ being the valence of ions, elementary electronic charge, average diffusivity of ions, bulk ionic number density and Boltzmann constant respectively. Besides, for a Newtonian fluid, the viscous dissipation term $\left(Q_{vd}\right)$ in equation (5) can be written as $\mu\left[2\left\{\left(\frac{\partial u}{\partial x}\right)^2+\left(\frac{\partial v}{\partial y}\right)^2\right\}+\left(\frac{\partial u}{\partial y}+\frac{\partial v}{\partial x}\right)^2\right]-\frac{2}{3}\mu\left(\nabla\cdot v\right)^2$ where $\mu(T)$ is the viscosity of the fluid. Here, we have neglected the variation of $\rho$ and $C_p$ with temperature in obtaining the temperature distribution, whereas other thermo-physical properties like viscosity ($\mu$), electrical permittivity ($\varepsilon$) and thermal conductivity ($k$) are considered to be temperature-dependent. The following forms of temperature dependences are assumed: $\mu = \mu_{ref} \exp\left[-\omega_1 \left(T - T_{ref}\right)\right]$, $\varepsilon = \varepsilon_{ref}\left[-\omega_2 \left(T - T_{ref}\right)\right]$ and $k = k_{ref} \exp\left[\omega_3 \left(T - T_{ref}\right)\right]$ where the subscript "*ref*" denotes reference value of property evaluated at reference temperature $T_{ref}$ with $\omega_i$s being individual temperature sensitivities. The reason for assuming constant $\rho$ and $C_p$ is that the relative change of $\rho$ and $C_p$ with temperature is insignificant as compared to the change of other parameters ($\mu$, $\varepsilon$, $k$) (Dietzel & Hardt 2017). For incompressible flow, $\nabla \cdot \boldsymbol{v}$ becomes zero and the expression of $Q_{vd}$ gets simplified. Now, to obtain the temperature field, we have non-dimensionalized the energy equation by using the following variables

$$\bar{u} = \frac{u}{u_c},\ \bar{v} = \frac{v\, l}{u_c\, h},\ \theta = \frac{T - T_C}{\Delta T_{ref}}$$

where $u_c$ and $T_C = T_{ref}$ are taken as characteristic scales of velocity and temperature with $\Delta T_{ref} = \left(T_H - T_C\right)/2$ being the characteristic temperature difference. The dimensionless form of equation (5) reads

$$\left.\begin{aligned} \beta\, Pe_T\left(\bar{u}\frac{\partial \theta}{\partial \bar{x}} + \bar{v}\frac{\partial \theta}{\partial \bar{y}}\right) = \beta^2 \frac{\partial}{\partial \bar{x}}\left(\frac{k}{k_{ref}}\frac{\partial \theta}{\partial \bar{x}}\right) + \frac{\partial}{\partial \bar{y}}\left(\frac{k}{k_{ref}}\frac{\partial \theta}{\partial \bar{y}}\right) + \beta^2 \frac{\varepsilon\, \kappa^2 D}{k_{ref}\, \Delta T_{ref}}\left(\frac{k_B T_C}{z e}\right)^2 \left(\frac{\partial \bar{\phi}}{\partial \bar{x}}\right)^2 \\ + \frac{\mu\, u_c^{\,2}}{k_{ref}\, \Delta T_{ref}}\left[2\beta^2\left\{\left(\frac{\partial \bar{u}}{\partial \bar{x}}\right)^2 + \left(\frac{\partial \bar{v}}{\partial \bar{y}}\right)^2\right\} + \left(\frac{\partial \bar{u}}{\partial \bar{y}} + \beta^2 \frac{\partial \bar{v}}{\partial \bar{x}}\right)^2\right] \end{aligned}\right\} \quad (6)$$

In equation (6), $\beta = h\, /\, l$ is the aspect ratio of the microchannel, $Pe_T = u_c\, h/\alpha_{ref}$ thermal Peclet number with thermal diffusivity $\alpha_{ref} = k_{ref}/\rho\, C_p$ and $\kappa = \sqrt{2\, z^2 e^2 n_\infty/\varepsilon\, k_B\, T}$ being the inverse of the electrical double layer (EDL) thickness. Similarly, temperature-dependent thermo-physical properties are also rewritten as $\bar{\mu} = \mu/\mu_{ref} = \exp\left(-\gamma C_\mu\, \theta\right)$, $\bar{\varepsilon} = \varepsilon/\varepsilon_{ref} = \exp\left(-\gamma C_\varepsilon\, \theta\right)$ and $\bar{k} = k/k_{ref} = \exp\left(\gamma C_k\, \theta\right)$ respectively where $C_\mu = \omega_1 T_C$, $C_\varepsilon = \omega_2 T_C$, $C_k = \omega_3 T_C$ and $\gamma = \Delta T / T_C$ is the ratio of the imposed temperature difference to reference temperature. Here, the values of $l$ and $h$ are taken as ~ 1 mm and 1 $\mu$m respectively thus making $\beta$ ~ O($10^{-3}$), i.e. << 1. The characteristic velocity scale is chosen as $u_c$ ~ $2\beta\, h\, n_\infty k_B\, T_c/\mu_{ref}$. Assuming viscosity, $\mu_{ref}$ ~ $10^{-3}$ Pa.s. and 1 mM concentration of electrolyte solution (i.e. $n_\infty = 6.023\times10^{23}$ $\text{mol}^{-1}$), $u_c$ turns out to be of the order of ~ $10^{-3}$ $ms^{-1}$. For typical values of parameters $\rho$ ~ $10^3$ kg $m^{-3}$, $C_p$ ~ 4200 J $kg^{-1}$ $K^{-1}$, $k_{ref}$ ~ 0.6 W $m^{-1}$ $K^{-1}$, the value of thermal Peclet number $\left(Pe_T\right)$ becomes ~ O($10^{-3}$). Since $Pe_T$ is already multiplied by another small quantity $\beta$, thus, one can safely neglect the advective component in the energy equation and therefore, any alteration in the hydrodynamics due to the application of thermal gradient has no effect on the temperature distribution. Similarly, heat generation term involves a quantity $\varepsilon\kappa^2 D k_B{}^2 T_c{}^2 / \left(k_{ref}\Delta T_{ref}\, z^2 e^2\right)$ which, for $\Delta T_{ref}$ ~ O(10) K, $\varepsilon$ ~ O($10^{-10}$) $Fm^{-1}$, $D$ ~ $10^{-9}$ $m^2s^{-1}$, $\kappa$ ~ O($10^7$) $m^{-1}$ becomes O($10^{-9}$) which is further multiplied by the quantity $\beta^2$ where $\beta$ << 1 and hence, the heat generation due to streaming field can be neglected. Also, most of the terms in viscous dissipation component involve $\beta^2$ and diminish for $\beta$ << 1 . The remaining term is $\dfrac{\mu\, u_c{}^2}{k_{ref}\,\Delta T_{ref}}\left(\dfrac{\partial\bar{u}}{\partial\bar{y}}\right)^2$, which in our analysis comes out to be O($10^{-4}$) and remains insignificant in determining the temperature field.

Case (a): *Temperature gradient applied in the axial direction*

For axially applied temperature gradient, the simplified energy equation subjected to the aforesaid assumptions is given below

$$\frac{\partial}{\partial\bar{x}}\left(\bar{k}\frac{\partial\theta}{\partial\bar{x}}\right)=0 \qquad (7)$$

where the dimensionless form of thermal conductivity is written as $\bar{k}=k/k_{ref}=\exp\left(\gamma C_k\,\theta\right)$ with $k_{ref}$ being reference thermal conductivity at $T_{ref}$. We have obtained both closed form and approximate analytical solution of the temperature distribution. For approximate analytical solution, well-known asymptotic approach has been followed (typically employed to capture small perturbation to the system) where any variable $\varphi$ can be expanded in the following way

$$\varphi=\varphi_0+\gamma\varphi_1+\gamma^2\varphi_2+........ \qquad (8)$$

where $\gamma=\Delta T/T_C$ is the thermal perturbation parameter used in the regular perturbation approach (8) which implies the ratio of the imposed temperature difference of the domain to the cold side temperature (so, $\gamma\to 0$ represents the scenario of isothermal condition). The exact solution of equation (7) subjected to axial temperature difference (i.e. at $\bar{x}=0, \theta=0$ and at $\bar{x}=1, \theta=2$) results

$$\theta = \ln\left\{-C_k C_{T1} \gamma \left(C_{T2} + x\right)\right\} / \gamma C_k \tag{9}$$

where $C_{T1} = -\left(e^{2\gamma C_k} - 1\right) / \gamma C_k$, $C_{T2} = 1 / \left(e^{2\gamma C_k} - 1\right)$. Meanwhile, the governing equations and the solution for asymptotic approach read as

$$\text{governing equations:} \quad \begin{aligned} &O(1): \ \frac{\partial^2 \theta_0}{\partial \bar{x}^2} = 0 && \text{(leading order)} \\ &O(\gamma): \ \frac{\partial^2 \theta_1}{\partial \bar{x}^2} + C_k \, \theta_0 \frac{\partial^2 \theta_0}{\partial \bar{x}^2} + C_k \left(\frac{\partial \theta_0}{\partial \bar{x}}\right)^2 = 0 && \text{(first order)} \end{aligned} \tag{10}$$

and solution:

$$\theta = \theta_0 + \gamma \theta_1 = 2\bar{x} + 2C_k \, \gamma\left(\bar{x} - \bar{x}^2\right) \tag{11}$$

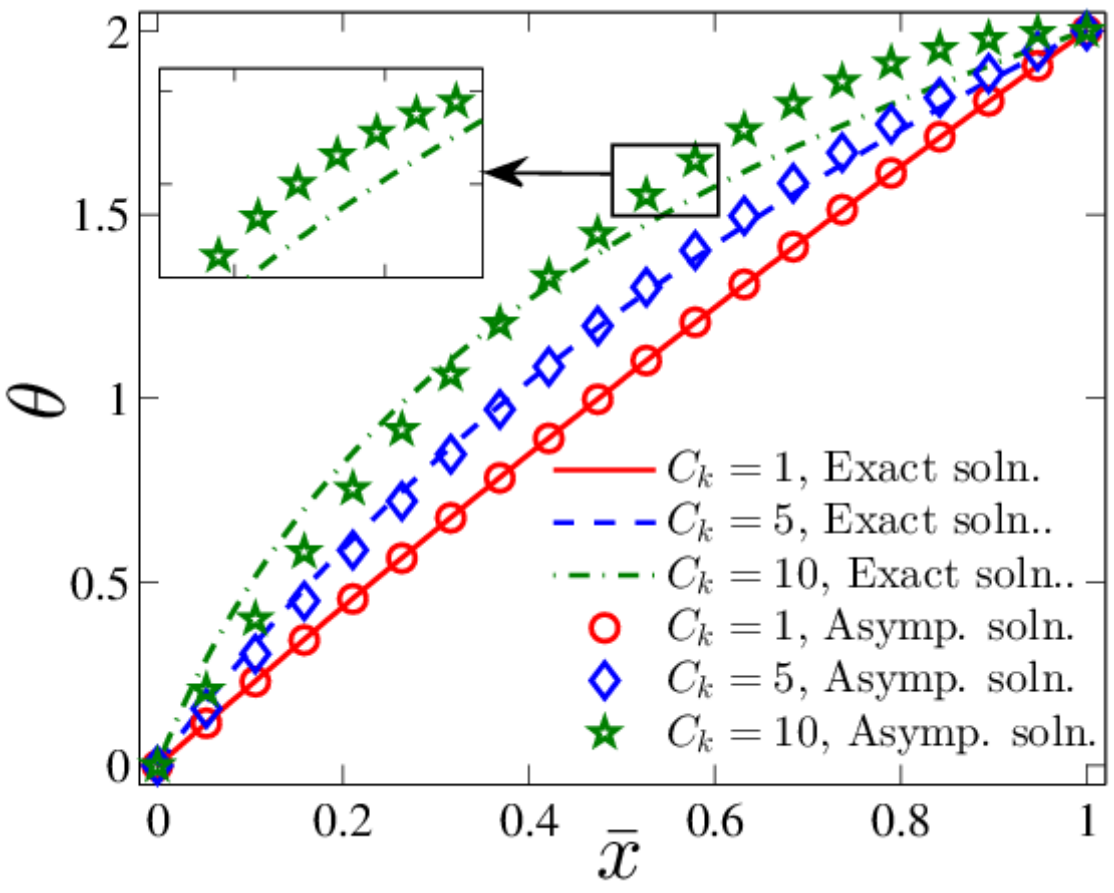

FIGURE 2. Dimensionless temperature profile in the *x*-direction for different $C_k$. Lines show exact solutions while asymptotic solutions are shown by symbols. Inset shows zoomed view at $C_k = 10$.

The comparison between the exact solution and asymptotic solution for temperature distribution is presented in figure 2 where the lines show the results of exact solution with symbols representing predictions of asymptotic approach. With increasing $C_k$, enhanced temperature sensitivity of the fluid thermal conductivity results in a departure from the linear variation of temperature in the axial direction. At higher $C_k$ (i.e. at $C_k$ = 10), this distribution becomes parabolic in nature where a deviation between the asymptotic and exact solution can be noticed (as shown in the inset). However, as reported in the literature, the relative change of thermal conductivity with temperature in typical electrolyte solution is $(1/k)\cdot(\partial k/\partial T) = 2.41\times10^{-3}$ $K^{-1}$ (Dietzel & Hardt 2017) for which the value of $C_k$ turns out to be of the order of unity and hence, a linear temperature distribution can be assumed safely as an approximation. Throughout our analysis, we maintain $C_k$ = 1 while presenting the results.

## 2.3 *Potential distribution*

In contrast to the conventional electrokinetics problem, in presence of thermal gradient, ions no longer remain in equilibrium and one cannot consider Boltzmann distribution assumption while obtaining the potential distribution. Here, we need to find the ionic number concentration first which can be obtained by employing classical Nernst-Planck (NP) equation. The electrolyte solution is considered to be dilute such that effects like ion-ion correlation or finite-size effect (also known as steric effect) can be neglected. Under steady state and in absence of any chemical reaction, the NP equation for transport of ionic species reads as $\nabla \cdot J_i = 0$ which means that the divergence of net ionic flux is zero. This ionic flux $\left(J_i\right)$ consists of four components, namely, advection $\left(n_i \boldsymbol{v}\right)$, diffusion $\left(D_i \nabla n_i\right)$, thermo-diffusion $\left(n_i D_{Ti} \nabla T\right)$ and electro-migration $\left(n_i u_i^* \nabla \phi\right)$ components

$$J_i = n_i \boldsymbol{v} - D_i \nabla n_i - n_i D_{Ti} \nabla T - n_i \mu_i^* \nabla \phi \tag{12}$$

In equation (12), $D_{Ti}$ and $\mu_i^*$ are the thermophoretic and electrophoretic mobilities respectively with $\mu_i^* = e z_i D_i / \left(k_B T\right)$. Using $\bar{n}_i = n_i / n_0$, $\bar{\phi} = z e \phi / \left(k_B T_C\right)$, $\bar{z}_i = z_i / z$, dimensionless form of $\nabla \cdot J_i = 0$ is written below

$$\beta^2 Pe_i \left( \bar{u} \frac{\partial \bar{n}_i}{\partial \bar{x}} + \bar{v} \frac{\partial \bar{n}_i}{\partial \bar{y}} \right) = \beta^2 \frac{\partial}{\partial \bar{x}} \left[ \frac{D_i}{D} \left( \frac{\partial \bar{n}_i}{\partial \bar{x}} + \bar{n}_i \bar{S}_{Ti} \gamma \frac{\partial \theta}{\partial \bar{x}} + \frac{\bar{z}_i \bar{n}_i}{1+\gamma\theta} \frac{\partial \bar{\phi}}{\partial \bar{x}} \right) \right] + \frac{\partial}{\partial \bar{y}} \left[ \frac{D_i}{D} \left( \frac{\partial \bar{n}_i}{\partial \bar{y}} + \bar{n}_i \bar{S}_{Ti} \gamma \frac{\partial \theta}{\partial \bar{y}} + \frac{\bar{z}_i \bar{n}_i}{1+\gamma\theta} \frac{\partial \bar{\phi}}{\partial \bar{y}} \right) \right] \tag{13}$$

Here, $\bar{S}_{Ti}$ is Soret coefficient of ions defined as $\bar{S}_{Ti} = \left(D_{Ti} / D_i\right) T_C$ and $Pe_i$ is ionic Peclet number $\left(Pe_i = u_c\, l / D\right)$ which should not exceed unity for a diffusion-dominated problem like this and therefore, advective term (i.e. left side of equation (13)) becomes ~ O ($\beta^2$). Since, $\beta << 1$, this term and the first term on the right side can be neglected while finding $\bar{n}_i$. Potential $\bar{\phi}$ consists of two terms $\bar{\phi} = \bar{\phi}\left(\bar{x}\right) + \psi\left(\bar{x}, \bar{y}\right)$ where $\bar{\phi}\left(\bar{x}\right)$ is the induced streaming field with $\bar{\psi}\left(x, y\right)$ being the potential induced within EDL. Considering this, the distribution of $\bar{n}_i$ subjected to symmetry condition at the channel centreline (i.e., at $\bar{y}$ = 0, $\partial \bar{n}_i / \partial \bar{y} = 0$) and number density being equal to bulk number concentration in electroneutral region ($\bar{n}_i = \bar{n}_{i\infty}$ at $\bar{\psi} = 0$) results

$$\overline{n}_i = \overline{n}_{i\infty} \exp\left(-\frac{\overline{\psi}}{1+\gamma\theta}\right) \tag{14}$$

where $\overline{\psi} = ze\psi/(k_B T_C)$. Since this shows an exponential dependence, it may seem like similar to the well-known Boltzmann distribution. However, the bulk ionic concentration $(\overline{n}_{i\infty})$ is not constant, instead it is varying axially in presence of the axial temperature gradient. To obtain this dependence, one need to equate the first term of right side of equation (13) with zero, so that

$$\frac{\partial \overline{n}_i}{\partial \overline{x}} + \overline{n}_i \, \overline{S}_{Ti} \, \gamma \frac{\partial \theta}{\partial \overline{x}} + \frac{\overline{z}_i \, \overline{n}_i}{1+\gamma\theta} \frac{\partial \overline{\phi}}{\partial \overline{x}} = 0 \tag{15}$$

Since electro-neutrality is prevalent in the bulk (i.e. $\partial\overline{\phi}/\partial\overline{x} = 0$), equation (15) gets simplified and we obtained the following expression for $\overline{n}_{i\infty}$

$$\overline{n}_{i\infty} = \exp\left(-\overline{S}_{Tavg} \, \gamma\theta\right) \tag{16}$$

The ionic number concentration described by the equations (14)-(16) is further used in the Poisson equation to evaluate the potential distribution which reads as $\nabla \cdot (\varepsilon \nabla \phi) = -\rho_e = -e\sum_i z_i \, n_i$. The dimensionless form of Poisson equation is given below

$$\overline{\varepsilon} \frac{\partial^2 \overline{\psi}}{\partial \overline{y}^2} = \overline{\kappa}_{eff}^2 \sinh\left(\frac{\overline{\psi}}{1+\gamma\theta}\right) \tag{17}$$

Contrary to the traditional electrokinetic studies, here EDL thickness (i.e. inverse of $\overline{\kappa}_{eff}$) no longer remains constant and becomes a function of the axial co-ordinate $\overline{\kappa}_{eff}^2 = \overline{\kappa}_0^2 \exp\left(-\overline{S}_{Tavg} \, \gamma\theta\right)$ which yields $\overline{\varepsilon} \frac{\partial^2 \overline{\psi}}{\partial \overline{y}^2} = \overline{\kappa}_0^2 \sinh\left[\overline{\psi}/(1+\gamma\theta)\right] \exp\left(-\overline{S}_{Tavg} \, \gamma\theta\right)$. For small values of surface potential and $\gamma$ (i.e. small imposed temperature difference), we can use Debye-Hückel linearization where $\sinh\left[\overline{\psi}/(1+\gamma\theta)\right]$ is approximated as $\overline{\psi}/(1+\gamma\theta)$; along with the exponential term being linearized as $\exp\left(-\overline{S}_{Tavg} \, \gamma\theta\right) \approx 1 - \overline{S}_{Tavg} \, \gamma\theta + \left(\overline{S}_{Tavg} \, \gamma\theta\right)^2/2$.

$$\left.\begin{aligned} &O(1): \ \frac{\partial^2 \overline{\psi}_0}{\partial \overline{y}^2} = \overline{\kappa}_0^2 \, \overline{\psi}_0 && (\text{leading order}) \\ &O(\gamma): \ \frac{\partial^2 \overline{\psi}_1}{\partial \overline{y}^2} - C_\varepsilon \theta_0 \frac{\partial^2 \overline{\psi}_0}{\partial \overline{y}^2} = \overline{\kappa}_0^2 \left(\overline{\psi}_1 - \overline{\psi}_0 \, \theta_0 - \overline{\psi}_0 \, \overline{S}_{Tavg} \, \theta_0\right) && (\text{first order}) \end{aligned}\right\} \tag{18}$$

In this context, it is worth mentioning that in presence of a thermal gradient, the surface potential (or zeta-potential) no longer remains constant. Instead, it becomes a strong function of the imposed temperature difference by following a linear relationship. Some previous experimental studies have demonstrated this temperature-dependence of zeta potential where it not only depends on temperature but also other factors like surface reactions (Ghonge *et al.* 2013; Ishido & Mizutani 1981; Reppert 2003; Revil *et al.* 2003). However, this surface reaction has relatively weaker influence on altering the non-isothermal flow dynamics, as highlighted in a recent study by Dietzel et al. (Dietzel & Hardt 2016). Consideration of this surface equilibrium and other associated reaction kinetics will unnecessarily complicate the present theoretical framework and hence, only linear dependence of zeta-potential with temperature is considered, i.e. at $\bar{y} = \pm 1$, $\bar{\psi} = \psi/\zeta = \left[1 + C_\zeta \, \gamma\theta\right]$ with $C_\zeta$ being the temperature sensitivity of zeta potential. Equation (18) subjected to the above boundary condition results in the following two equations for the potential distribution

$$\bar{\psi}_0 = \frac{\cosh\left(\bar{\kappa}_0 \, \bar{y}\right)}{\cosh\left(\bar{\kappa}_0\right)} \tag{19}$$

$$\bar{\psi}_1 = f_1 \exp\left(\bar{\kappa}_0 \, \bar{y}\right) + f_2 \exp\left(-\bar{\kappa}_0 \, \bar{y}\right) - \frac{C_{\varepsilon eff}\left\{\exp\left(-\bar{\kappa}_0 \, \bar{y}\right) - 4\,\bar{\kappa}_0 \, \bar{y} \sinh\left(\bar{\kappa}_0 \, \bar{y}\right) + 2\cosh\left(\bar{\kappa}_0 \, \bar{y}\right)\right\}}{8\cosh\left(\bar{\kappa}_0\right)} \tag{20}$$

The coefficients of equation (20) are shown in Section A1 of the supplementary material where the modified potential distribution for temperature-dependent of zeta potential is also presented. Reported experimental results of temperature dependence of zeta potential (Reppert 2003) is depicted in figure 3a (i) which clearly shows that zeta potential ($\zeta$) is indeed strongly dependent on temperature by following a linear relationship. These data are fitted in the form of $\zeta/\zeta_{ref} = m\left(T - T_{ref}\right)$ (this is the same form of zeta potential variation with temperature which we present as $\bar{\psi} = \left[1 + C_\zeta \, \gamma\theta\right]$) and this approximates quite well with the experimental data. As can be seen from this figure, depending on electrolyte concentration, this sensitivity with temperature is increased almost twice as the slope ($m$) of fitted line increases from 0.0112 to 0.0214. In the dimensionless form, the value of $C_\zeta$ turns out to be varying from ~ 3 to 6 (approximately). Accordingly, in our analysis, we have chosen $C_\zeta$ to lie between 0 and 4 while presenting the results, with $C_\zeta$ = 0 representing the case of constant zeta potential. Figure 3a (ii) shows the potential distribution in the transverse direction where the inset clearly shows the effect of $C_\zeta$ on the potential distribution. Effect of $C_\zeta$ is noticeable only in close vicinity of channel walls (where EDL is located) while remains ineffective in the electro-neutral region. As $C_\zeta$ is increased, more is the ionic redistribution within EDL and the magnitude of zeta potential may increase up to twice (at $\gamma = 0.1$ as shown in the inset) as observed in figure 3a (iii) for $C_\zeta = 4$.

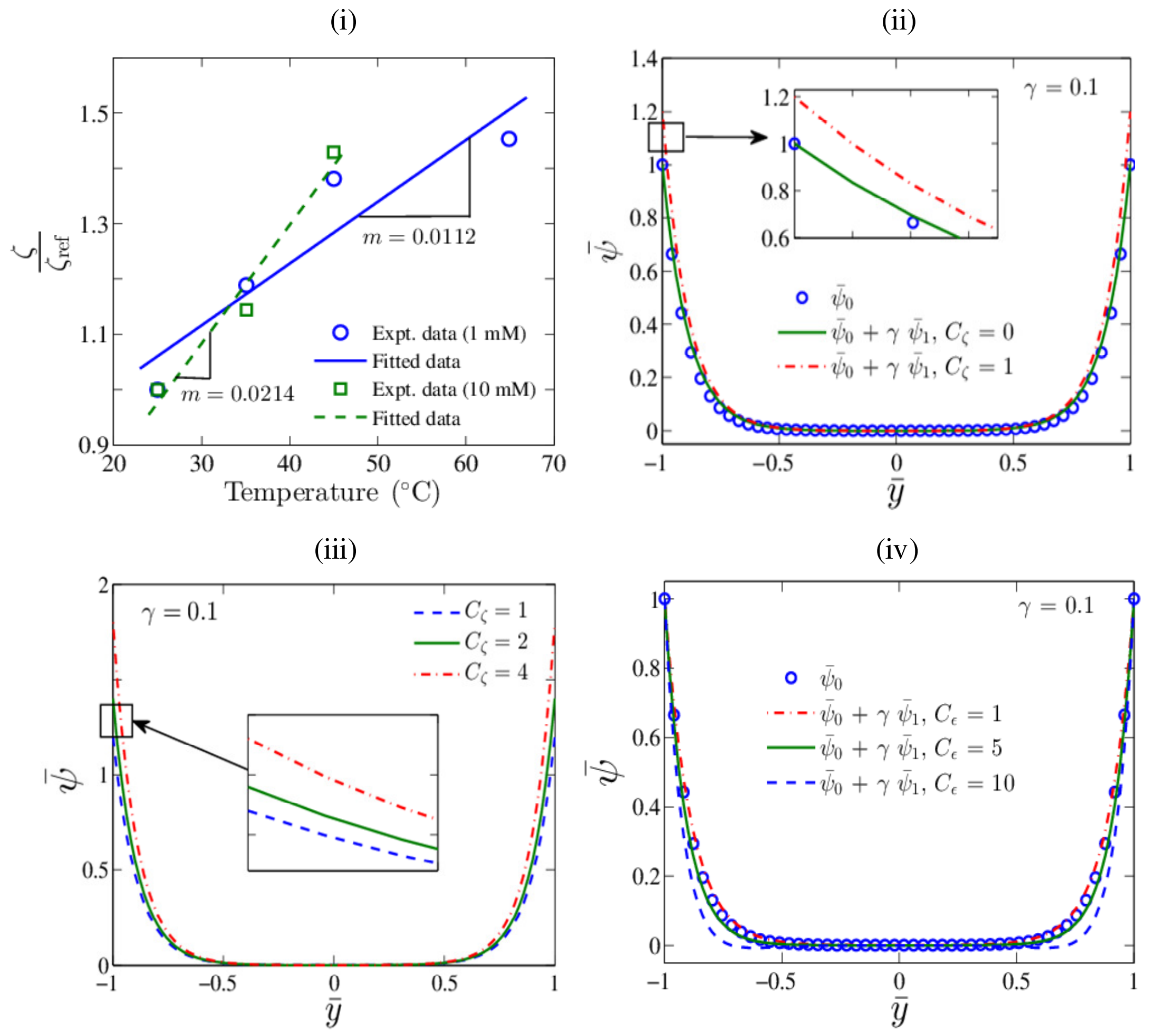

FIGURE 3a. (i) Reported experimental results of zeta potential variation with temperature (Reppert 2003), (ii) potential distribution in the *y*-direction for both constant and temperature-dependent zeta potential, variation of the same for (iii) different $C_\zeta$ and for (iv) different $C_\varepsilon$.

Another key factor in altering the potential distribution is $C_\varepsilon$ representing the change in electrical permittivity with temperature. Increasing $C_\varepsilon$ from 1 to 10 indicates significant reduction in electrical permittivity whose effect is reflected in the charge distribution via the permittivity-induced component $C_\varepsilon\,\theta_0\left(\partial^2\bar{\psi}_0/\partial\bar{y}^2\right)$. As observed in figure 3a (iv), increasing $C_\varepsilon$ creates a deviation in potential profile close to the channel walls while remains unaffected in the bulk.

### 2.4 *Velocity distribution*

Because of the application of the externally imposed temperature gradient ($\Delta T$), a thermo-electric field is induced within the microchannel by virtue of the difference in Soret coefficients between cations and anions. Besides, one can expect another form of thermoelectricity which can be induced due to the difference in ionic diffusivities of the ions even if their Soret coefficients remain same. Also, since the thermo-physical properties of the fluid are strongly dependent on temperature, there will be drastic alteration in the hydrodynamics in presence of $\Delta T$. For example, the variation of electrical permittivity with temperature is manifested through the contribution of the dielectrophoretic body force with concomitant induction of axial pressure gradient thereby strongly influencing the advection motion of fluid. For steady, incompressible flow, the velocity distribution for a Newtonian fluid is described by continuity equation $\nabla \cdot \boldsymbol{v} = 0$ along the following momentum equation

$$0 = -\nabla p + \nabla \cdot \mu \left[ \nabla \boldsymbol{v} + \left( \nabla \boldsymbol{v} \right)^T \right] + F_{EK} + F_{DEP} \tag{21}$$

The left hand side of equation (21) is zero because of negligible inertial effect consideration which occurs only when the Reynolds number (*Re*) associated with flow is very less compared to unity, i.e. *Re* << 1. Here, $p$ is the hydrodynamic pressure, $\mu \left[ \nabla \boldsymbol{v} + \left( \nabla \boldsymbol{v} \right)^T \right]$ viscous stress, $F_{EK} = -\rho_e \nabla \phi$ electrokinetic force and $F_{DEP} = -\left(1/2\right)\left(\nabla \phi\right)^2 \nabla \varepsilon$ dielectrophoretic force respectively. In the momentum equation, the effect of electrostriction force (i.e. the variation of electrical permittivity with fluid density) is not considered because of the assumption of constant density of the fluid. Evaluating $F_{EK}$ and $F_{DEP}$ from the previously-obtained charge distribution, substituting those in momentum equation and employing $\beta$ << 1, we get

$$\left.\begin{aligned} x-\text{component}:\ 0 &= -\frac{\partial \bar{p}}{\partial \bar{x}} + \bar{\mu}\frac{\partial^2 \bar{u}}{\partial \bar{y}^2} + \frac{\lambda \bar{\varepsilon}}{\bar{\kappa}_{eff}^2}\left[ \frac{\partial^2 \bar{\psi}}{\partial \bar{y}^2}\frac{\partial \bar{\phi}}{\partial \bar{x}} - \frac{1}{2}\gamma \Delta T_{ref}\left( \frac{\partial \bar{\psi}}{\partial \bar{y}} \right)^2 \frac{\partial \theta}{\partial \bar{x}} \right] \\ y-\text{component}:\ 0 &= -\frac{\partial \bar{p}}{\partial \bar{y}} + \frac{\lambda \bar{\varepsilon}}{\bar{\kappa}_{eff}^2}\frac{\partial^2 \bar{\psi}}{\partial \bar{y}^2}\frac{\partial \bar{\psi}}{\partial \bar{y}} \end{aligned}\right\} \tag{22}$$

In equation (22), $\lambda$ represents the ratio of induced velocity due to osmotic pressure to the characteristic flow velocity $\lambda = 2\beta\, h\, n_\infty k_B T_c / \mu_{ref} u_c$ . First, we solve the pressure distribution from the $y$-component of the momentum equation. Here, the axial pressure-gradient consists of two terms, one is the externally imposed pressure gradient while the other part is induced in non-isothermal condition. Using this, the modified form of the $x$-component of the momentum equation reads as

$$\bar{\mu}\frac{\partial^2 \bar{u}}{\partial \bar{y}^2} = \frac{\partial \bar{p}_0}{\partial \bar{x}} + \frac{\lambda}{\bar{\kappa}_0^2}\exp\left(\bar{S}_{Tavg}\gamma\theta\right)\left\{\bar{\varepsilon}\frac{\partial^2 \bar{\psi}}{\partial \bar{y}^2}\bar{E}_x + \frac{1}{2}\frac{\partial \bar{\varepsilon}}{\partial \bar{x}}\left(\frac{\partial \bar{\psi}}{\partial \bar{y}}\right)^2\right\} + \frac{\lambda}{2}\frac{\partial \theta}{\partial \bar{x}}\frac{\partial_\theta \bar{n}_\infty}{\bar{n}_\infty}\frac{\bar{\psi}^2}{(1+\gamma\theta)} - \frac{\lambda}{2}\frac{\partial \theta}{\partial \bar{x}}\gamma\frac{\bar{\psi}^2}{(1+\gamma\theta)^2} \quad (23)$$

where $\bar{E}_x = -\partial\bar{\phi}/\partial\bar{x}$ . Equation (23) is now solved by following the aforesaid asymptotic approach where the governing equations at different orders of perturbations are as follows

$$\left.\begin{aligned}
&O(1): \quad \frac{\partial^2 \bar{u}_0}{\partial \bar{y}^2} = \frac{\partial \bar{p}_0}{\partial \bar{x}} + \frac{\lambda}{\bar{\kappa}_0^2}\frac{\partial^2 \bar{\psi}_0}{\partial \bar{y}^2}\bar{E}_{x0}, \\
&O(\gamma): \frac{\partial^2 \bar{u}_1}{\partial \bar{y}^2} - C_\mu\,\theta_0\frac{\partial^2 \bar{u}_0}{\partial \bar{y}^2} = \frac{\lambda}{\bar{\kappa}_0^2}\begin{bmatrix}\dfrac{\partial^2 \bar{\psi}_0}{\partial \bar{y}^2}\left\{\bar{E}_{x0}\,\theta_0\left(\bar{S}_{Tavg} - C_\varepsilon\right) + \bar{E}_{x1}\right\} \\ \dfrac{\partial^2 \bar{\psi}_1}{\partial \bar{y}^2}\bar{E}_{x0} - \dfrac{1}{2}C_\varepsilon\dfrac{\partial \theta_0}{\partial \bar{x}}\left(\dfrac{\partial \bar{\psi}_0}{\partial \bar{y}}\right)^2\end{bmatrix} - \frac{\lambda}{2}\bar{\psi}_0^2\frac{\partial \theta_0}{\partial \bar{x}}\left(1 + \bar{S}_{Tavg}\right)
\end{aligned}\right\} \quad (24)$$

To obtain the flow field, equation (24) is subjected to no-slip condition at the surface $(\bar{u} = 0 \text{ at } \bar{y} = 1)$ and symmetry condition $(\partial\bar{u}/\partial\bar{y} = 0 \text{ at } \bar{y} = 0)$ at the channel centreline. Using this, the velocity distribution are given in the following two equations

$$\bar{u}_0 = \frac{1}{2}\frac{\partial \bar{p}_0}{\partial \bar{x}}\left(\bar{y}^2 - 1\right) + \frac{\lambda}{\bar{\kappa}_0^2}\bar{E}_{x0}\left[\frac{\cosh\left(\bar{\kappa}_0\,\bar{y}\right)}{\cosh\left(\bar{\kappa}_0\right)} - 1\right] \quad (25)$$

$$\begin{aligned}
\bar{u}_1 = {} & \beta_2\frac{\partial \bar{p}_0}{\partial \bar{x}}\frac{\bar{y}^2}{2} + \beta_2\frac{\lambda}{\bar{\kappa}_0^2}\bar{E}_{x0}\frac{\cosh\left(\bar{\kappa}_0\,\bar{y}\right)}{\cosh\left(\bar{\kappa}_0\right)} + \beta_3\left(\beta_1\bar{E}_{x0} + \bar{E}_{x1}\right)\frac{\cosh\left(\bar{\kappa}_0\,\bar{y}\right)}{\cosh\left(\bar{\kappa}_0\right)} - \frac{\beta_5\,\bar{\kappa}_0^2\,F_1\left(\bar{y}\right) + \beta_6\,F_2\left(\bar{y}\right)}{\bar{\kappa}_0\cosh^2\left(\bar{\kappa}_0\right)} \\
& + \beta_4\,\bar{E}_{x0}\left[\frac{\beta_7}{\bar{\kappa}_0^2}\exp\left(\bar{\kappa}_0\,\bar{y}\right) + \frac{\left(\beta_8 - \beta_{11}\right)}{\bar{\kappa}_0^2}\exp\left(-\bar{\kappa}_0\,\bar{y}\right) + \beta_9\,F_3\left(\bar{y}\right) + \beta_{10}\,F_4\left(\bar{y}\right)\right] + c_1\,\bar{y} + c_2
\end{aligned} \quad (26)$$

Equation (25) represents the perturbation-free flow field which is the well-known expression for combined pressure-driven and streaming field induced electrokinetic flow while equation (26) indicates the contribution of the thermal perturbation to the flow field. The coefficients of equation (26) are given in Section A2 of the supplementary material. Still, the completion of the flow field requires the knowledge of the induced streaming potential for which the electroneutrality condition (i.e. vanishing net current condition) is invoked, i.e. $I_{\text{net}} = I_{\text{streaming}} + I_{\text{advection}} = 0$. For sake of conciseness, the governing equations for streaming field evaluation are presented in Section A3 of the supplementary material while the final forms of $\overline{E}_x$'s are shown below

$$\overline{E}_{x0} = \frac{2\,Pe_i \dfrac{\partial \overline{p}_0}{\partial \overline{x}} \left\{ \tanh\left(\overline{\kappa}_0\right) - \overline{\kappa}_0 \right\}}{\left[ 2\,\overline{\kappa}_0^3 + \overline{\kappa}_0^2 \left\{ \dfrac{\overline{\kappa}_0}{2\cosh^2\left(\overline{\kappa}_0\right)} + \dfrac{1}{2}\tanh\left(\overline{\kappa}_0\right) \right\} - 2\chi\,\overline{\kappa}_0^2 \tanh\left(\overline{\kappa}_0\right) - Pe_i\,\lambda \left\{ \dfrac{\overline{\kappa}_0}{\cosh^2\left(\overline{\kappa}_0\right)} - \tanh\left(\overline{\kappa}_0\right) \right\} \right]} \quad (27)$$

and
$$\overline{E}_{x1} = \left( -\gamma_{48} \frac{\partial \theta_0}{\partial \overline{x}} - Pe_i\,\gamma_{47} - \overline{E}_{x0}\,\gamma_{45} \right) \Big/ \gamma_{46} \quad (28)$$

The coefficients of equation (28) can be found in Section A4 of the supplementary material. Here, it is necessary to mention about the two parameters involved in the final expression of the streaming potential described by equations (27)-(28), $\chi$ and $C_D$. $\chi$ represents the molecular diffusivity difference between the co-ions and counter-ions $\chi = \left(D_+ - D_-\right) / \left(D_+ + D_-\right)$ while $C_D$ represents the sensitivity of diffusivity of ions with temperature (here diffusivity $D$ is assumed to be obeying the following relationship $D = D_{ref}\left[1 + \omega_4\left(T - T_{ref}\right)\right]$, i.e. $\overline{D} = D/D_{ref} = \left[1 + C_D\,\gamma\theta\right]$ (D. R. Lide 2005; Ghonge *et al.* 2013)). While results for varying $\chi$ are reported in the main paper, corresponding results for $C_D$ are presented in the supplementary material (Section A5) for the sake of brevity. Once the velocity field is known, we obtain the volumetric flow rate $\left(\overline{Q}\right)$ through the microchannel by integrating the flow velocity cross-sectionally

$$\overline{Q} = \overline{Q}_0 + \gamma\overline{Q}_1 = \int_{-1}^{1} \overline{u}\,d\overline{y} = \int_{-1}^{1} \left(\overline{u}_0 + \gamma\overline{u}_1\right) d\overline{y} \quad (29)$$

Similarly, the dispersion coefficient can also be expressed as

$$\overline{D}_{eff} = \overline{D}_0 + \overline{D}_1 \quad (30)$$

where the estimation of $\overline{D}_1$ (i.e. dispersion coefficient at O($\gamma$)) involves the knowledge of O($\gamma$) flow velocity $\overline{u}_1$.

Case (b): *Temperature gradient applied in the transverse direction*

For temperature difference applied in the transverse direction, considering all previously-mentioned assumptions and simplifications (for the case of axially applied thermal gradient), the governing equation of temperature field reads as

$$\frac{\partial}{\partial \bar{y}}\left(\bar{k}\frac{\partial \theta}{\partial \bar{y}}\right)=0 \tag{31}$$

and the corresponding temperature profile (asymptotic solution) is given by

$$\theta=\theta_0+\gamma\theta_1=\left(\bar{y}+1\right)+\frac{\gamma C_k}{2}\left(1-\bar{y}^2\right) \tag{32}$$

Now, the governing equation for the transport of ionic species reads as

$$\frac{\partial}{\partial \bar{y}}\left\{\frac{D}{D^*}\left(\frac{\partial \bar{n}_i}{\partial \bar{y}}+\bar{S}_{Ti}\,\gamma\frac{\partial \theta}{\partial \bar{y}}\bar{n}_i+\frac{\bar{n}_i\,\bar{z}_i}{1+\gamma\theta}\frac{\partial \bar{\psi}}{\partial \bar{y}}\right)\right\}=0 \tag{33}$$

which results the following distribution of the ionic number concentration within the EDL

$$\bar{n}_+-\bar{n}_-\approx-\Delta\bar{S}_T\,\gamma\theta-\frac{2\bar{\psi}}{\left(1+\gamma\theta\right)}-2\gamma\frac{\partial \theta}{\partial y}\int\frac{\bar{\psi}}{\left(1+\gamma\theta\right)^2}d\bar{y} \tag{34}$$

This distribution is now used in the Poisson equation to obtain the potential distribution. The governing equations for the potential distribution are as follows

$$\left.\begin{aligned}&\mathrm{O}(1):\ \frac{\partial^2\bar{\psi}_0}{\partial \bar{y}^2}=\bar{\kappa}_0^2\,\bar{\psi}_0\\&\mathrm{O}(\gamma):\ \frac{\partial^2\bar{\psi}_1}{\partial \bar{y}^2}-\bar{\kappa}_0^2\,\bar{\psi}_1=\left\{C_\varepsilon-\left(1+\bar{S}_{Tavg}\right)\right\}\bar{\kappa}_0^2\,\theta_0\,\bar{\psi}_0+C_\varepsilon\frac{\partial \theta_0}{\partial \bar{y}}\frac{\partial \bar{\psi}_0}{\partial \bar{y}}+\bar{\kappa}_0^2\frac{\partial \theta_0}{\partial \bar{y}}\int\bar{\psi}_0\,d\bar{y}+\frac{\bar{\kappa}_0^2\,\Delta\bar{S}_T\,\theta_0}{2}\end{aligned}\right\} \tag{35}$$

For potential distribution, unlike the case of axial temperature gradient, no closed form solution is possible. So, one needs to either employ the asymptotic approach (approximate solution) or can solve it numerically.

$$\mathrm{O}(1):\qquad \bar{\psi}_0=\frac{\cosh\left(\bar{\kappa}_0\,\bar{y}\right)}{\cosh\left(\bar{\kappa}_0\right)}$$

$$\mathrm{O}(\gamma):\ \bar{\psi}_1=\left[\begin{aligned}&\left\{-\frac{f_1(\bar{y})\cosh\left(\bar{\kappa}_0\,\bar{y}\right)}{8\,\bar{\kappa}_0\cosh\left(\bar{\kappa}_0\right)}-\frac{\Delta\bar{S}_T\,f_3(\bar{y})}{4\,\bar{\kappa}_0}\right\}g_1(\bar{y})\exp(-\bar{\kappa}_0\,\bar{y})+c_1\exp(-\bar{\kappa}_0\,\bar{y})+c_2\exp(\bar{\kappa}_0\,\bar{y})\\&+\left\{-\frac{f_2(\bar{y})\,g_2(\bar{y})}{16\,\bar{\kappa}_0\cosh\left(\bar{\kappa}_0\right)}+\frac{\Delta\bar{S}_T\,f_4(\bar{y})\,g_3(\bar{y})}{4\,\bar{\kappa}_0\cosh\left(\bar{\kappa}_0\right)}\right\}\exp\left(\bar{\kappa}_0\,\bar{y}\right)-\frac{\left\{g_4(\bar{y})\exp(\bar{\kappa}_0\,\bar{y})-g_5(\bar{y})\exp(-\bar{\kappa}_0\,\bar{y})\right\}}{8\cosh\left(\bar{\kappa}_0\right)}\end{aligned}\right] \tag{36}$$

The coefficients of equation (36) are reported in Section B1 of the supplementary material.

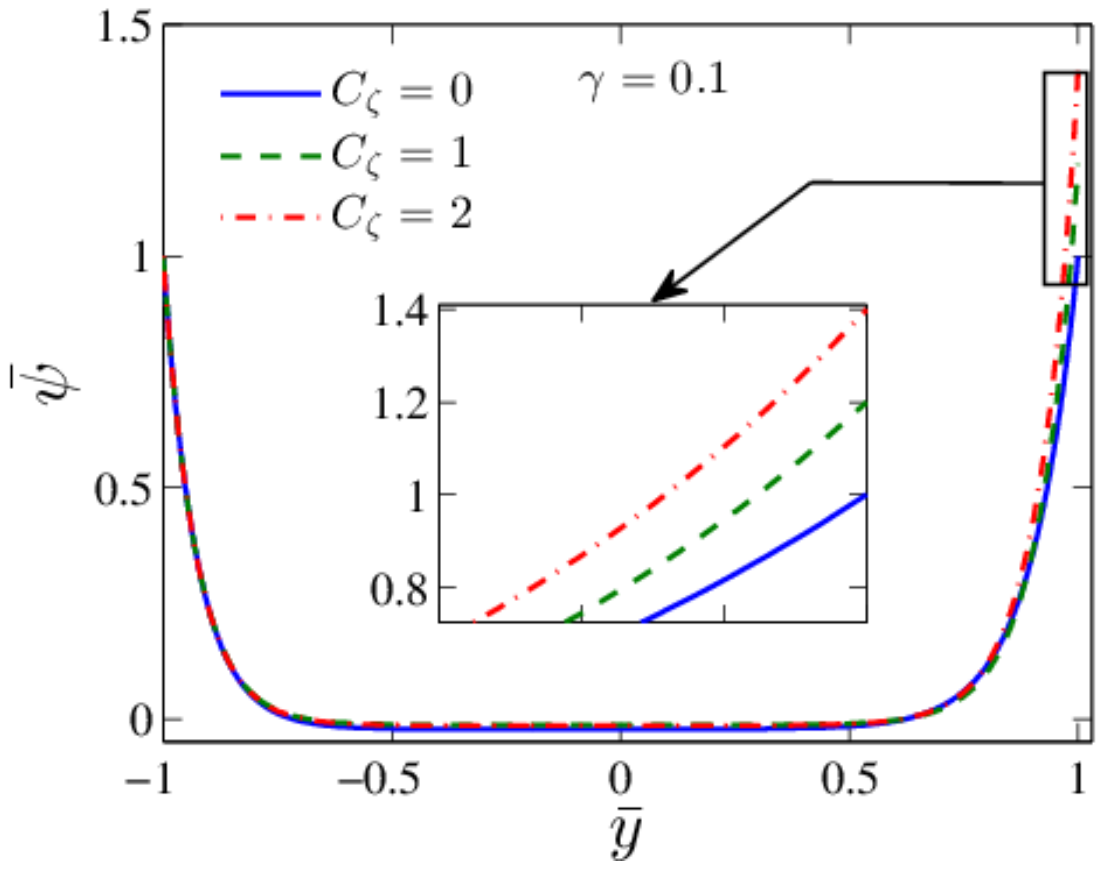

FIGURE 3b. Potential distribution in the $y$-direction (at $\gamma$ = 0.1) for transverse temperature gradient. Inset shows the zoomed view towards the top wall (i.e. close to $\bar{y}$ = 1).

Comparing the potential distribution profiles for longitudinal and transverse thermal gradients one can understand that the basic difference is the introduction of an asymmetry through the temperature distribution which has its immediate effect on the potential distribution within the EDL. With increasing $C_\zeta$, the magnitude of surface potential gets amplified significantly in the hot region while remaining unaffected in the cold region thus creating more asymmetry in the $y$-direction. This further influences the fluid advective motion through electrokinetic forcing thereby altering strongly the velocity distribution. Knowing the potential distribution, the flow field can now be obtained from the following governing equations

$$\mathrm{O}(1): \quad \frac{\partial^2 \bar{u}_0}{\partial \bar{y}^2} = \frac{\partial \bar{p}_0}{\partial \bar{x}} + \frac{\lambda}{\bar{\kappa}_0^2} \bar{E}_{x0} \frac{\partial^2 \bar{\psi}_0}{\partial \bar{y}^2} \tag{37}$$

and
$$\mathrm{O}(\gamma): \quad \frac{\partial^2 \bar{u}_1}{\partial \bar{y}^2} - C_\mu \theta_0 \frac{\partial^2 \bar{u}_0}{\partial \bar{y}^2} - C_\mu \frac{\partial \theta_0}{\partial \bar{y}} \frac{\partial \bar{u}_0}{\partial \bar{y}} = \frac{\lambda}{\bar{\kappa}_0^2} \left[ \begin{array}{l} \dfrac{\partial^2 \bar{\psi}_0}{\partial \bar{y}^2} \bar{E}_{x0} \theta_0 \left( \bar{S}_{Tavg} - C_\varepsilon \right) + \dfrac{\partial^2 \bar{\psi}_0}{\partial \bar{y}^2} \bar{E}_{x1} \\ + \dfrac{\partial^2 \bar{\psi}_1}{\partial \bar{y}^2} \bar{E}_{x0} - \dfrac{\partial \theta_0}{\partial \bar{y}} C_\varepsilon \dfrac{\partial \bar{\psi}_0}{\partial \bar{y}} \bar{E}_{x0} \end{array} \right] \tag{38}$$

The solution of these two equations subjected to no-slip boundary condition at the surfaces yields

$$\mathrm{O}(1): \bar{u}_0 = \frac{1}{2} \frac{\partial \bar{p}_0}{\partial \bar{x}} \left( \bar{y}^2 - 1 \right) + \frac{\lambda}{\bar{\kappa}_0^2} \bar{E}_{x0} \left[ \frac{\cosh\left( \bar{\kappa}_0 \bar{y} \right)}{\cosh\left( \bar{\kappa}_0 \right)} - 1 \right] \tag{39}$$

$$\left.\begin{aligned} O(\gamma):\bar{u}_1 = C_\mu \left[\frac{\partial \bar{p}_0}{\partial \bar{x}}\left(\frac{\bar{y}^3}{3}+\frac{\bar{y}^2}{2}\right)+\frac{\lambda \bar{E}_{x0} f_7(\bar{y})}{\cosh(\bar{\kappa}_0)}\right]-\frac{C_\varepsilon \lambda \bar{E}_{x0} f_7(\bar{y})}{\cosh(\bar{\kappa}_0)}+\frac{\bar{E}_{x1}\lambda \cosh(\bar{\kappa}_0 \bar{y})}{\bar{\kappa}_0^2 \cosh(\bar{\kappa}_0)}+\frac{\bar{S}_{Tavg}\lambda \bar{E}_{x0} f_6(\bar{y})}{\cosh(\bar{\kappa}_0)} \\ +\frac{\lambda \bar{E}_{x0}\Delta \bar{S}_T}{2}\left(\frac{\bar{y}^3}{3}+\frac{\bar{y}^2}{2}\right)+\frac{\lambda \bar{E}_{x0}(C_\varepsilon+1)}{\bar{\kappa}_0^2}\frac{\sinh(\bar{\kappa}_0 \bar{y})}{\cosh(\bar{\kappa}_0)}-\frac{\lambda \bar{E}_{x0} f_6(\bar{y}) f_5}{\cosh(\bar{\kappa}_0)}+C_{u1}\bar{y}+\lambda \bar{E}_{x0} I_6 + C \end{aligned}\right\} \quad (40)$$

The coefficients of equation (40) are presented in Section B2 of the supplementary material. Now, the induced streaming field ($\bar{E}_x$) can be evaluated by using the electroneutrality condition similarly as mentioned in the case of axial temperature gradient (governing equations for electroneutrality condition are shown in Section B3 of the supplementary material)

$$\bar{E}_{x0} = \frac{2Pe_i \dfrac{\partial \bar{p}_0}{\partial \bar{x}}\left\{\bar{\kappa}_0 - \tanh(\bar{\kappa}_0)\right\}}{\left[Pe_i \lambda \left\{\dfrac{\bar{\kappa}_0}{\cosh^2(\bar{\kappa}_0)} - \tanh(\bar{\kappa}_0)\right\} + 2\,\bar{\kappa}_0^2\,\chi \tanh(\bar{\kappa}_0)\right]} \quad (41)$$

and

$$\bar{E}_{x1} = -\left(\alpha_{32} + \alpha_{30}\bar{E}_{x0}\right)/\alpha_{31} \quad (42)$$

with the coefficients shown in Section B3 of the supplementary material.

### 2.5 *Fluid rheology*

Now, we focus on the alteration of hydrodynamics and associated streaming field caused by the inclusion of rheological aspects of fluid. Towards this, the constitutive form of viscoelastic fluid has been chosen for which we have taken into account the simplified Phan-Thien Tanner (sPTT) model, typically employed to model the rheological characteristics of viscoelastic fluids. Here, the basic difference with Newtonian fluid lies in the expressions of the stress tensors (Afonso *et al.* 2009; Arcos *et al.* 2018; Bautista *et al.* 2013; Mukherjee *et al.* 2017a, 2017b)

$$\left.\begin{aligned} 2\mu_{eff}\frac{\partial u}{\partial x} &= F\tau_{xx} + \lambda_{eff}\left(u\frac{\partial \tau_{xx}}{\partial x} + v\frac{\partial \tau_{xx}}{\partial y} - 2\frac{\partial u}{\partial x}\tau_{xx} - 2\frac{\partial u}{\partial y}\tau_{yx}\right) \\ 2\mu_{eff}\frac{\partial v}{\partial y} &= F\tau_{yy} + \lambda_{eff}\left(u\frac{\partial \tau_{yy}}{\partial x} + v\frac{\partial \tau_{yy}}{\partial y} - 2\frac{\partial v}{\partial x}\tau_{xy} - 2\frac{\partial v}{\partial y}\tau_{yy}\right) \\ \mu_{eff}\left(\frac{\partial u}{\partial y} + \frac{\partial v}{\partial x}\right) &= F\tau_{xy} + \lambda_{eff}\left(u\frac{\partial \tau_{xy}}{\partial x} + v\frac{\partial \tau_{xy}}{\partial y} - \frac{\partial u}{\partial y}\tau_{yy} - \frac{\partial v}{\partial x}\tau_{xx}\right) \end{aligned}\right\} \quad (43)$$

One important thing to note here that, in presence of temperature gradient, not only fluid viscosity but also fluid relaxation time start to become temperature-dependent and interestingly, the extent of viscoelasticity of a fluid is governed by these two parameters - viscosity and relaxation time. In equation (43), $F$ is the stress coefficient function, which for linear PTT model, takes the form $F = 1 + \delta \lambda_{eff} \left( \tau_{xx} + \tau_{yy} \right) / \mu_{eff}$ where $\delta$ represents the extensibility of viscoelastic fluid (here $\delta$ is chosen to be equal to unity). Here, $\mu_{eff}$ and $\lambda_{eff}$ are viscosity and relaxation time of the fluid which are assumed to be function of temperature as $\mu_{eff} = \mu_{ref} \exp\left[ -\omega_1 \left( T - T_{ref} \right) \right]$ and $\lambda_{eff} = \lambda_{ref} \exp\left[ -\omega_4 \left( T - T_{ref} \right) \right]$ respectively. The dimensionless form of fluid relaxation time is $\bar{\lambda} = \lambda_{eff} / \lambda_{ref} = \exp\left( -\gamma C_\lambda \theta \right)$ with $C_\lambda$ being the sensitivity of relaxation time with temperature. Unlike the Newtonian fluid, evaluation of the leading order streaming potential $\left( \bar{E}_{x0} \right)$ in case of a viscoelastic fluid involves a cubic equation in the form of $A \bar{E}_{x0}^{\,3} + B \bar{E}_{x0}^{\,2} + C \bar{E}_{x0} + D = 0$ (the expressions for *A*, *B*, *C* and *D* can be found in Section C1 of the supplementary material) in which the real root has been chosen for further calculations while the other two roots are complex conjugate to each other and thus discarded.

One important non-dimensional number associated with the flow of viscoelastic fluids is Deborah number (*De*) which determines the degree of viscoelasticity of a fluid defined as $De = \lambda_{ref} \kappa_{ref} u_c$. Keeping other parameters ($\kappa_{ref}, u_c$) fixed, the value of Deborah number (*De*) can lie in between $10^{-1}$ and 1 depending on fluid relaxation time ($\lambda_{ref}$) (which may vary due to factors like polymer concentration, polymer molecular weight etc.). Accordingly, we have chosen *De* in our analysis to vary in between 0 to 1 (with *De* = 0 representing the results of a Newtonian fluid) while obtaining the results. Meanwhile, the stress-tensors in case of viscoelastic fluid are made dimensionless as follows

$$\bar{\tau}_{xx} = \frac{\tau_{xx} h}{\mu_{ref} u_c}, \bar{\tau}_{yy} = \frac{\tau_{yy} h}{\mu_{ref} u_c}, \bar{\tau}_{xy} = \frac{\tau_{xy} h}{\mu_{ref} u_c}$$

The governing equations related to viscoelastic fluids for both axial and transverse thermal gradient can be found in Section C2 of the supplementary material.

## 3. Results and discussions

Since this analysis involves large set of parameters, numerous results can be obtained by combining all pertinent parameters. However, for improved readability, we have highlighted some key results involving velocity distribution, induced streaming potential, volumetric flow rate, and finally the dispersion coefficient (which in turn depends on previous parameters). For representing the results, we have employed the following ranges of the involved parameters: $0 \leq \gamma$ (thermal perturbation parameter) $\leq 0.1$, $10^{-1} \leq C_\mu$ (sensitivity coefficient of viscosity) $\leq 10$, $10^{-1} \leq C_\varepsilon$ (sensitivity coefficient of electrical permittivity) $\leq 10$, $C_k$ (sensitivity coefficient of thermal conductivity) = 1, $-0.3 \leq \chi$ (diffusivity difference between ions) $\leq 0$, $0 \leq \Delta\bar{S}_T$ (difference in Soret coefficients between ions) $\leq 1$, $1 \leq \bar{\kappa}_0$ (degree of confinement) $\leq 10$, $0 \leq -\partial\bar{p}_0/\partial\bar{x}$ (strength of imposed pressure gradient) $\leq 10$, $0 \leq C_D$ (sensitivity coefficient of diffusivity of ions) $\leq 5$, $0 \leq C_\zeta$ (sensitivity coefficient of zeta potential) $\leq 4$ and $0 \leq De$ (Deborah number) $\leq 1$.

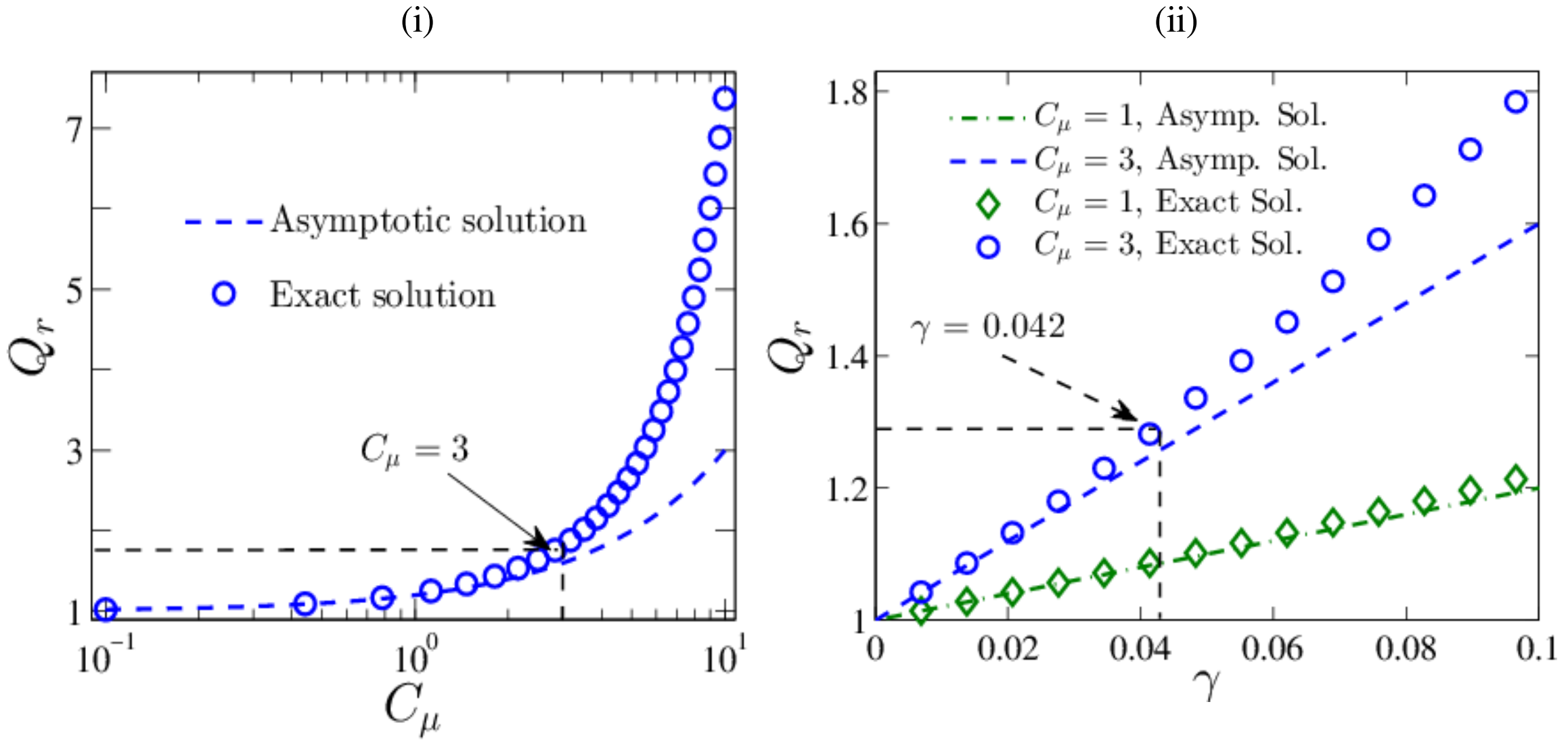

FIGURE 4. Volume flow rate ratio with (i) $C_\mu$ and (ii) $\gamma$ respectively. Symbols show the results obtained from exact solution while lines are for asymptotic solution.

First we have shown the validity of range of some parameters involved in our analysis. As already mentioned, both approximate and exact solutions are obtained for axially applied thermal gradient. Figure 4(i) depicts the comparison of volumetric flow rates between the asymptotic (approximate) and exact solutions as a function of $C_\mu$ which denotes the sensitivity of fluid viscosity with temperature. Here, $C_\mu$ is varied over two decades ranging from $10^{-1}$ to $10^{1}$ while keeping other parameters constant. As evident, asymptotic solution closely approximates the exact solution up to $C_\mu = 3$. Beyond this critical value of $C_\mu$, large deviation between these two solutions is observed with asymptotic approach under-predicting the results significantly at higher $C_\mu$. In figure 4(ii), comparison of the same has been shown for varying $\gamma$, which denotes the ratio of the imposed temperature difference with respect to the reference temperature, i.e. a quantification of degree of thermal perturbation imposed. In this context, it is necessary to note that our asymptotic analysis is corrected up to first order in $\gamma$. Therefore, our findings are able to capture the linear thermal effect only. For lower $C_\mu$, the asymptotic solution approximates the exact solution reasonably well while at higher $C_\mu$, deviation takes place between the two solutions with asymptotic solution under-predicting volumetric flow rate beyond $\gamma = 0.042$. In the similar way, comparisons between asymptotic and numerical solutions have been made for the case of transverse temperature gradient which for the sake of conciseness are presented in Section D of the supplementary material.

### 3.1 *Effect of axial temperature gradient*

In figure 5(a), the variation of streaming potential ratio ($E_r$) is plotted against some key parameters. Here, streaming potential ratio ($E_r$) is defined as the ratio of the streaming potential induced due to the combined action of externally imposed thermal and pressure gradient to that induced due to the sole action of pressure gradient. While realizing the alteration in the flow field upon thermal gradient, one can expect the effect of the electrical permittivity variation induced dielectrophorteic force on the flow field by inducing an axial pressure gradient, i.e. osmotic pressure gradient due to excess charge redistribution. Also, further source of alteration is expected through the physical property variation where these properties depend strongly on the temperature distribution thereby influencing strongly the fluid motion. However, the description of the flow physics is not completed here because one needs to look into the inherent temperature dependence of the ionic species. This contribution comes into picture via the ionic species transport, typically known as Soret effect. The movement of the ionic species in thermal gradient is a response subjected to the imposed temperature gradient. This effect is incorported through the thermodiffusion term $\left(n_i\, D_{Ti}\nabla T\right)$ in the Nernst-Planck equation where the Soret coefficient $\left(S_{Ti} = D_{Ti}/D_i = Q_i/k_B T^2\right)$ depends on the heat of transport of ions ($Q_i$). Here, $Q_i$ is the quantification of the degree of sensitivity of ionic mobility with temperature, i.e. thermophoretic mobility of ions. Now, let us first consider the sole action of Soret effect on the flow dynamics by assuming the absence of any other pertinent forces. Positive value of $Q_i$ suggests us that ions should have a propensity of moving towards the cold region from the hot region and as a result a thermo-electric field should be induced in the same direction while role-revesral should be observed for negative $Q_i$. Interestingly, the conventional streaming field arising from the imposed pressure gradient is induced in the reverse direction to the applied pressure gradient, i.e. from the hot region to the cold region. So, the thermal gradient induced thermo-electric streaming field (assuming positive value of $Q_i$) seems to act in the same direction of the pressure gradient induced streaming field and intuition tells us that the combined action of these two gradients should result in an enhancement of the net induced streaming potential. However, the situation does not remain same if there is a difference in thermophoretic mobilities between the co-ions (here negative ions) and counter-ions (positive ions). The extent of thermophobic behavior, i.e. the tendency of ions moving away from the hot region can change depending on the difference in the thermophoretic mobilities ($\Delta\overline{S}_T$) between cations and anions. Increasing $\Delta\overline{S}_T$ indicates higher heat of transport of counter-ions compared to co-ions, so the counter-ions are more likely to move towards the cold region than the co-ions. This leads to a clear axial separation between the ions and gives rise to an accumulation of the counter-ions in the upstream section, i.e. in the cold region. This induces a form of thermo-electric field towards the downstream (because of $\Delta\overline{S}_T$) while the another form of it is formed towards the upstream (because of the sole action of Soret effect). Hence, the net streaming field (due to combined pressure-driven and temperature gradient induced flow) depends on the relative strength of these two counteracting thermo-electric fields. For very low value of $\Delta\overline{S}_T$, these two opposing factors are comparable to each other and the streaming potential ratio ($E_r$) is close to unity up to $\Delta\overline{S}_T$ = 0.05. As one starts increasing $\Delta\overline{S}_T$, the effect of $\Delta\overline{S}_T$-induced streaming field overshadows the temperature gradient induced streaming field and net streaming potential ratio experiences massive reduction. For higher $\Delta\overline{S}_T$, its effect becomes so pronounced that it completely nullifies the pressure-gradient-induced streaming field (shown by the red colored solid line in figure 5a (i)).

Now, we look into the expression of the bulk number density of ions which reads as $\left(\partial \ln \overline{n}_i / \partial \overline{x}\right) = -\overline{S}_{Ti}\left(\partial \theta / \partial \overline{x}\right)$ which clearly tells us that the application of the external $\Delta T$ in the positive $x$-direction induces an axial concentration gradient of ions in the opposite direction,.i.e. from the hot region to the cold region. The concentration gradient $\left(\partial \overline{n}_i / \partial \overline{x}\right)$ creates a migration of the ions towards the upstream section. Now, we focus our attention towards the definition of the parameter $\chi = \left(D_+ - D_-\right)/\left(D_+ + D_-\right)$ which is an indicator of the diffusivity difference between the cations and anions. Positive value of $\chi$ implies that the diffusivity of cations (here counter-ions) is higher compared to the anions which leads to more migration of counter-ions towards upstream. As a result, there will be an accumulation of counter-ions in the upstream while in the downstream there will be more co-ions. This segregation of ions in different axial locations creates an axial separation between them thus creating stronger induced streaming field, while negative value of $\chi$ means more mobility of anions thereby leads to more accumulation of co-ions in the upstream section resulting weaker streaming field. Here we have shown the variation of streaming potential ratio ($E_r$) for $\chi$ ranging from -0.2 to 0 (figure 5a (ii)). Here, the value of $\chi$ is ion-specific. For typical electrolyte solutions like aqueous KCl, NaCl, LiCl, the value of $\chi$ ranges between ~ -0.3 to 0 (approximately) (Zhang *et al.* 2019) and we have chosen the range of $\chi$ accordingly while presenting the results.

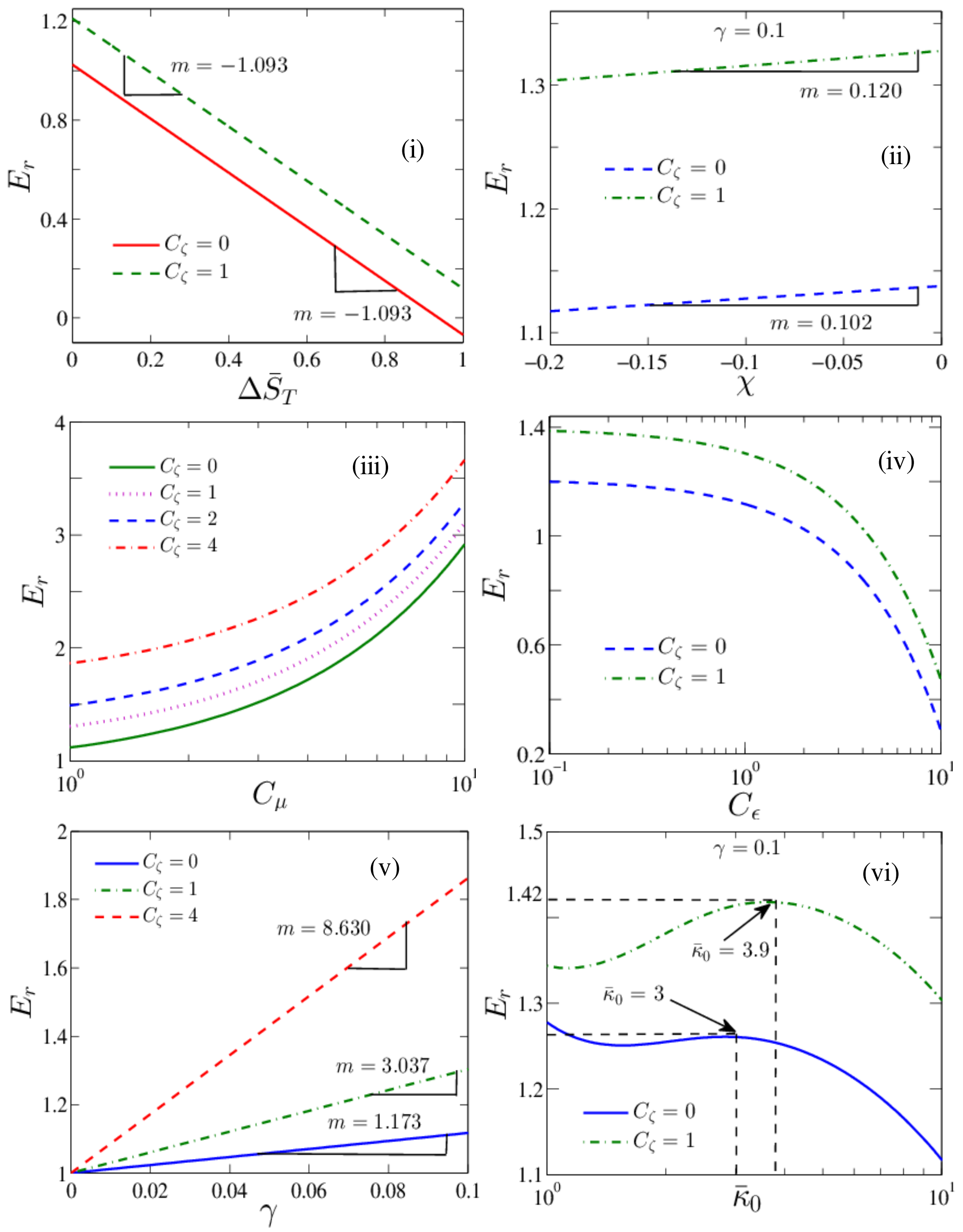

FIGURE 5a. Dependence of streaming potential ratio ($E_r$) with (i) $\Delta\bar{S}_T$, (ii) $\chi$, (iii) $C_\mu$, (iv) $C_\varepsilon$, (v) $\gamma$ and (vi) $\bar{\kappa}_0$ respectively.

On closely observing the governing equations involving velcoity profile and induced streaming field one can understand that the contribution of $C_\mu$ comes through the viscous resistance term in the fluid advective motion and the subsequent advective current calculation involved in the electroneutrality condition. Since $C_\mu$ indicates the temperature sensitivity of viscosity with temperature, increasing $C_\mu$ means viscosity becomes more susceptible to any change in temperature thus resulting strong reduction in fluid viscosity. Therefore, the viscous resistance to the flow reduces to a great extent. Thus, the induced streaming field due to migration of ions upon applied pressure gradient and the streaming field due to migration of ions upon induced concentration gradient assist each other (both are induced in the direction from the hot region to the cold region) resulting significant augmentation in the streaming potential ratio ($E_r$). Here, increasing $C_\mu$ from 1 to 10 results in ~ 3 times (shown by the green colored solid line) augmentation of the streaming potential ratio ($E_r$), as visible from figure 5a (iii).

Now, examining the equation describing the potential distribution (equation (17)), it can be observed that unlike the conventional electrokinetic problem, here the EDL thickness ($\lambda_D$) no longer remains constant. Instead, the effective EDL thickness ($\lambda_{Deff}$) becomes a strong function of temperature and departs significantly from its reference value (of isothermal condition) in the following way: $\bar{\lambda}_{D_{eff}} = \bar{\lambda}_{D_0}\sqrt{\exp\left\{-\gamma\theta\left(C_\varepsilon - \bar{S}_{Tavg}\right)\right\}}$. Keeping other parameters fixed, with increasing $C_\varepsilon$ (which is the temperature sensitivity of electrical permittivity) streaming potential ratio ($E_r$) decreases sharply by following an exponential thinning behavior. Since electrical permittivity decreases with temperature, the thickness of EDL also decreases which results lesser penetration of diffuse layer of EDL to the bulk, so more is the region of electroneutrality (which means the region where there is equal number of counter-ions and co-ions). Now, the strength of streaming current involved in the streaming potential estimation depends on this degree of penetration. In the region $10^{-1} \leq C_\varepsilon \leq 1$, the rate of reduction of $E_r$ with $C_\varepsilon$ is not significant, it decreases slightly from 1.2 to 1.1. However, this reduction gets amplified beyond $C_\varepsilon \sim 1$, $E_r$ decreases sharply and falls to ~ 0.2 at $C_\varepsilon = 10$ (figure 5a (iv)).

For fixed value of the aforesaid factors like $\Delta\bar{S}_T$, $\chi$, $C_\mu$, $C_\varepsilon$, since this analysis is restricted up to order $\gamma$, it is expected to get linear dependence of perturbation parameter on the flow field and streaming potential. For constant zeta potential ($C_\zeta = 0$), $E_r$ is increased up to ~ 1.12 times as $\gamma$ is varying from 0 to 0.1. However, for temperature-dependent zeta potential, its enhanced sensitivity with temperature results in significant increment of streaming potential where $E_r$ increases up to ~ 1.3 and ~ 1.85 times for $C_\zeta = 1$ and $C_\zeta = 4$ respectively (figure 5a (v)).

Now, the degree of confinement is incorporated within the parameter $\bar{\kappa}_0$, i.e. inverse of the thickness of EDL. The variation of streaming potential ratio ($E_r$) with $\bar{\kappa}_0$ is shown in figure 5a (vi). Increasing $\bar{\kappa}_0$ at one side increases the region of electroneutrality (ensuring a reduction of streaming current), while on the other hand, the strength of acting electrokinetic forces gets modulated and net $E_r$ is decided by these two factors. For constant zeta-potential case (i.e. $C_\zeta = 0$), $E_r$ first decreases with $\bar{\kappa}_0$ up to $\bar{\kappa}_0 = 1.7$, then increases slowly up to $\bar{\kappa}_0 = 3$ beyond which it falls sharply resulting only ~ 1.12 times increment of $E_r$ visible at higher $\bar{\kappa}_0$ ($\bar{\kappa}_0 = 10$). However, the presence of amplified zeta potential (for temperature-depedent zeta potential case, i.e. $C_\zeta = 1$), makes role reversal of $\bar{\kappa}_0$ as $E_r$ first increases with $\bar{\kappa}_0$ in the region $1 \leq \bar{\kappa}_0 \leq 3.9$, beyond which similar decaying behavior (as seen for $C_\zeta = 0$) with $\bar{\kappa}_0$ is observed (figure 5a (vi)).

Now we recall the boundary condition involved in evaluating the potential distribution, i.e. $\bar{\psi} = \left[1 + C_\zeta \, \gamma\theta\right]$. Instead of being constant, zeta potential gradually developes in the axial direction. Any alteration in the surface potential creates a perturbation in the near-wall fluid velocity and affects the adjacent layer of flow through viscous interaction. This axial variation of zeta potential affects the fluid momentum transport by generating a secondary component of flow. Overall, one important conclusion from figure 5(a) is that as far as the thermo-electrical energy conversion is concerned, the inclusion of temperature-dependent zeta potential should be necessary else this could lead to a grossly erroneous estimation of induced streaming potential.

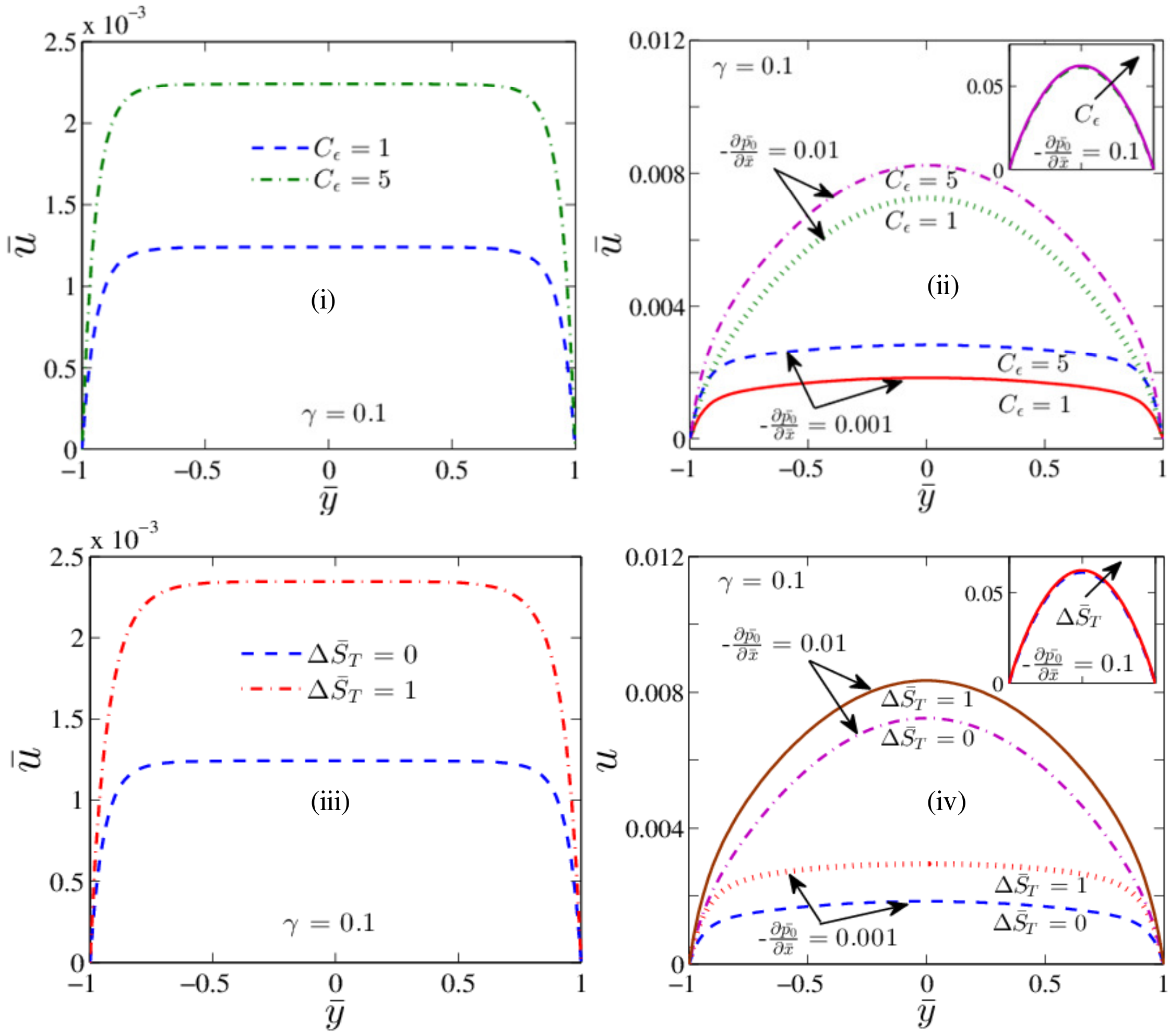

FIGURE 5b. Velocity profile in the *y*-direction (i) in absence and (ii) in presence of pressure gradient for different $C_\varepsilon$, variation of the same (iii) in absence and (iv) in presence of pressure gradient for different $\Delta\bar{S}_T$ (evaluated at $\bar{x} = 1$).

Figure 5(b) mainly highlights the effect of two parameters $C_\varepsilon$ and $\Delta\bar{S}_T$ on the velocity distribution both in the absence and presence of external pressure gradient. In absence of pressure gradient, the flow physics is solely governed by the external temperature difference and the velocity profile here follows uniform plug-type distribution (visible in both figures 5b (i) and (iii)) which is also typically observed in purely electroosmotic flows. As one starts introducing pressure gradient $\left(-\partial\bar{p}_0/\partial\bar{x} = 0.001\right)$, departure from uniformity in flow field is noticeable and at higher strength of pressure-gradient $\left(-\partial\bar{p}_0/\partial\bar{x} = 0.01\right)$, velocity distribution becomes parabolic similar to the Poiseuille flow. At higher strength, effect of pressure-gradient becomes so dominant that it dictates the flow physics where the effect of thermal gradient gets overshadowed. As already discussed, keeping other parameters constant, increasing $C_\varepsilon$ results in higher sensitivity of electrical permittivity with temperature which results in attenuation of the EDL thickness thus leads to a reduction in the streaming current and the induced streaming field. So, the strength of streaming-field-driven back electroosmotic flow also decreases thereby resulting an enhancement in the magnitude of the net flow velocity. In absence of pressure gradient, an increment of ~ 1.8 times can be observed as one increases $C_\varepsilon$ from 1 to 5. Now, the rate of increment in velocity magnitude upon increasing $C_\varepsilon$ gets dampened as one introduces pressure gradient where increment up to ~ 1.54 times and ~ 1.14 times in flow velocity are observed for $-\partial\overline{p}_0/\partial\overline{x} = 0.001$ and $-\partial\overline{p}_0/\partial\overline{x} = 0.01$ respectively as seen from figure 5b (ii). Increasing pressure gradient beyond 0.01 makes it so dominant that the effect of $C_\varepsilon$ becomes indistinguishable (inset of figure 5b (ii).

The effect of $\Delta\overline{S}_T$ on the velocity field is shown in figures 5b (iii)-(iv) where increasing $\Delta\overline{S}_T$ signifies increasing heat of transport of counter-ions creating enhanced axial separation of ions. The resulting thermo-electric streaming field acts in the opposite direction to that induced due to concentration gradient (which is induced due to external $\Delta T$) driven migration of ions thus leads to a suppression of the net streaming field and reverse electrokinetic flow. As a result, augmentation up to ~ 1.9 times in velocity magnitude is seen as $\Delta\overline{S}_T$ is varying from 0 to 1. Here also, with growing pressure gradient, its effect on the flow field starts to be important thus suppressing the effectiveness of $\Delta\overline{S}_T$. As clear from figure 5b (iv), as pressure gradient is increased 10 folds $\left(-\partial\overline{p}_0/\partial\overline{x} = 0.01\right)$, velocity magnitude with $\Delta\overline{S}_T$ is increased only up to ~ 1.15 times (while for $-\partial\overline{p}_0/\partial\overline{x} = 0.001$, this ratio is ~ 1.6). Beyond this, effect of $\Delta\overline{S}_T$ on the flow field becomes inconsequential (evident from the inset).

In traditional pressure-driven flow of electrolyte solution, the induced streaming field creates a flow in the reverse direction, thus leading to a suppression of the net volumetric flow rate of pressure-driven flow. Now, in presence of thermal gradient, net throughput through the channel depends on whether the induced thermo-electric streaming field assists or opposes the pressure-gradient induced streaming field. Here, we have presented some results of flow rate ratio ($Q_r$) which is defined as the ratio of the net flow due to combined action of external pressure gradient and temperature gradient to the throughput because of sole action of pressure gradient. For solely temperature gradient driven flow, as $\Delta\overline{S}_T$ is increasing, the reduction in net streaming potential results in significant enhancement in flow rate in absence of pressure gradient, .i.e. increment up to ~ 2.1 times can be observed (as shown in Figure 5c (i)). This rate gets attenuated in presence of pressure gradient (at $-\partial\overline{p}_0/\partial\overline{x} = 0.01$). One interesting thing to notice that there is a cross-over between the graphs for $-\partial\overline{p}_0/\partial\overline{x} = 0$ and $-\partial\overline{p}_0/\partial\overline{x} = 0.01$ at $\Delta\overline{S}_T$ = 0.66. Below this critical $\Delta\overline{S}_T$, $Q_r$ is higher in presence of pressure gradient and beyond this, reverse trend has been observed. In solely $\Delta T$ driven flow, at lower $\Delta\overline{S}_T$, axial separation between ions is not that higher thus creating lower reduction of streaming potential whereas at higher $\Delta\overline{S}_T$, this occurs relatively faster thus giving rise to more augmnetation in flow rate as compared to the case of $-\partial\overline{p}_0/\partial\overline{x} = 0.01$. Increasing $\Delta\overline{S}_T$ has almost vanishing effect in higher strength of pressure gradient. As shown in figure 5c (i) slight increase in flow rate ratio is seen for $-\partial\overline{p}_0/\partial\overline{x} = 0.1$ while for $-\partial\overline{p}_0/\partial\overline{x} = 1$, the profile of $Q_r$ remains constant at 1.2.

The variation of $Q_r$ with $\chi$ is shown in figure 5c (ii). Decreasing $\chi$ results in lowering the induced streaming potential, because of enhanced migration of co-ions in the upstream section. However, the dependence of $Q_r$ on $\chi$ is very weak as slight increment in flow rate ratio is observed when $\chi$ is decreased from 0 to -0.2. The magnitude of the flow rate ratio ($Q_r$) becomes higher (from 1 to ~ 1.55) as one introduces pressure gradient $\left(-\partial\overline{p}_0/\partial\overline{x} = 0.01\right)$. Now, increasing the strength of pressure gradient results in reduction of the magnitude of $Q_r$ with $\chi$ becoming increasingly insignificant ($Q_r$ vs. $\chi$ plot remains constant).

Now, the flow rate ratio ($Q_r$) dependence on $C_\mu$ is depicted in figure 5c (iii) where increasing $C_\mu$ denotes reduction in the fluid viscosity with temperature. The subsequent lowering in the viscous resistance of flow results significant augmentation of the volumetric flow rate. However, as described earlier in the variation of streaming potential ratio ($E_r$), increasing $C_\mu$ also ensures faster migration of the ions caused by the imposed pressure gradient or the induced concentration gradient which also generates more electrokinetic flow in the reverse direction. Therefore, the net throughput depends on the rate of lessening of viscous resistance as well as the enhanced induced streaming field. For lower strength of pressure gradient ($-\partial\overline{p}_0/\partial\overline{x} = 0.01$), an enhancement in the flow rate ratio ($Q_r$) up to ~ 3.35 times is observed as $C_\mu$ is increased from 1 to 10. Increasing the strength of pressure gradient lowers the magnitude of the net flow rate with the dependence on $C_\mu$ remaining qualitatively similar.

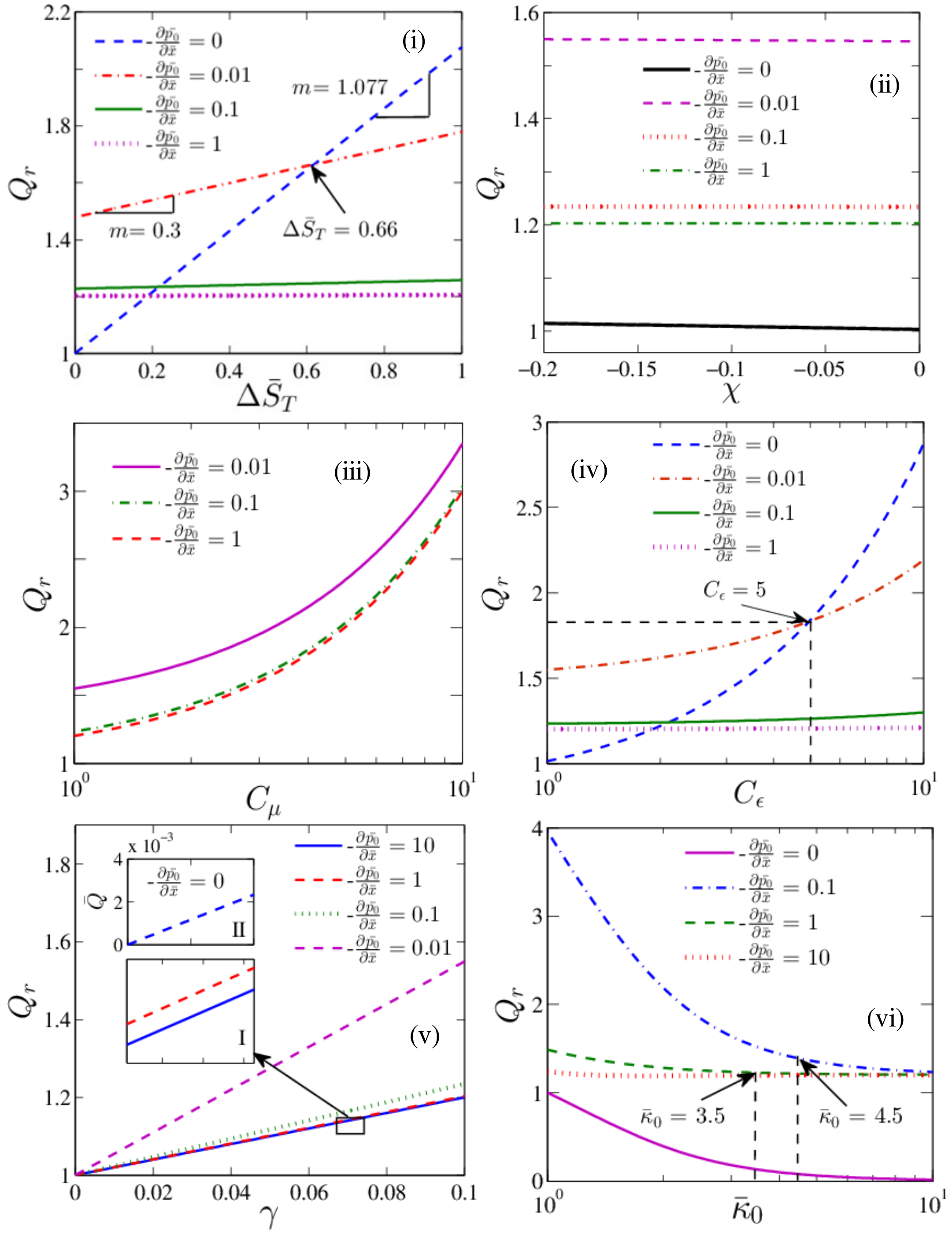

FIGURE 5c. Flow rate ratio variation with (i) $\Delta\bar{S}_T$, (ii) $\chi$, (iii) $C_\mu$, (iv) $C_\varepsilon$, (v) $\gamma$ and (vi) $\bar{\kappa}_0$ respectively (evaluated at $\bar{x} = 1$).

The alteration in the flow rate ratio ($Q_r$) with increasing $C_\varepsilon$ is highlighted in figure 5c (iv) where the reduction in the streaming potential owing to the temperature-sensitivity of electrical permittivity causes significant enhancement (~ 2.87 times) in the net throughput for purely thermally driven flow. However, the impact of $C_\varepsilon$ on $Q_r$ becomes less in presence of pressure gradient and interestingly, a cross-over between plots of $-\partial\overline{p}_0/\partial\overline{x}=0$ and $-\partial\overline{p}_0/\partial\overline{x}=0.1$ has been observed at $C_\varepsilon$ = 5. Higher the strength of pressure gradient, $C_\varepsilon$ becomes less important and the profile of $Q_r$ remains flat at a constant value of ~ 1.21.

The dependence of $Q_r$ on $\gamma$ is shown in figure 5c (v) where higher value of $\gamma$ implies higher degree of thermal perturbation which is more pronounced at lower strength of pressure gradient $\left(-\partial\overline{p}_0/\partial\overline{x}=0.01\right)$. This influence of $\gamma$ gets weakened with increasing pressure gradient. As clear from the figure, the slope of $Q_r$ vs. $\gamma$ becomes almost half of the previous case as the strength of pressure gradient increases 10 times from $-\partial\overline{p}_0/\partial\overline{x}=0.01$ to $-\partial\overline{p}_0/\partial\overline{x}=0.1$. At higher strength, the graphs are identical with zoomed view of a portion (inset I of figure 5c (v)) indicating negligible difference in $Q_r$ with $\gamma$. For completeness, inset II of figure 5c (v) shows the variation of dimensionless flow rate for purely $\Delta T$ driven flow which also shows linear dependence with increasing thermal perturbation.

Figure 5c (vi) shows the variation of flow rate ratio ($Q_r$) with decreasing channel confinement (i.e. increasing $\overline{\kappa}_0$). For purely thermally driven flow (i.e. $-\partial\overline{p}_0/\partial\overline{x}=0$), $Q_r$ decreases sharply from its reference value and approaches towards zero as one increases $\overline{\kappa}_0$ from 1 to 10. At higher $\overline{\kappa}_0$, the net flow through the channel gets completely arrested. This also highlights the importance of channel confinement on the hydrodynamics of purely thermally driven flow. As pressure gradient is introduced $\left(-\partial\overline{p}_0/\partial\overline{x}=0.1\right)$, the reduction in $Q_r$ becomes so rapid that it falls from 4 times to approach a constant value of ~ 1.2 beyond $\overline{\kappa}_0=4.5$. Now, increasing the strength of pressure gradient makes it predominant to dictate the flow physics. Therefore, the rate of reduction of $Q_r$ with $\overline{\kappa}_0$ gets dampened and approaches to the same constant value earlier (at $\overline{\kappa}_0=3.5$) as compared to the case of $-\partial\overline{p}_0/\partial\overline{x}=0.1$. Also, the plots for $-\partial\overline{p}_0/\partial\overline{x}=1$ and $-\partial\overline{p}_0/\partial\overline{x}=10$ becomes identical at higher $\overline{\kappa}_0$ indicating vanishing effect of $\overline{\kappa}_0$ on flow field.

Figure 5d highlights the dependence of the dispersion coefficient ratio ($\bar{D}_{eff}$) with the involving parameters where $\bar{D}_{eff}$ means the ratio of the net dispersion resulting from the combined action of two driving forces, pressure gradient and thermal gradient with respect to that arising from the sole action of pressure gradient. From the definition of hydrodynamic dispersion coefficient (please refer to equation (4)), it is clearly evident that any small change in the flow field results in introducing higher perturbation in the dispersion coefficient as compared to the net volumetric throughput. This is because, the estimation of dispersion coefficient involves parameters like non-dimensional average velocity and plate height where the plate height further depends on the square of the mean velocity thus showing strong dependence on the flow field. Contrary to the Poiseuille flow, in case of electroosmotic flow, because of the uniformity in flow field, main contribution of solute dispersion in absence of shear-induced dispersion comes from molecular diffusional dispersion thus resulting lower extent of dispersion. Since pressure-driven flow of electrolyte induces an electroosmotic flow in the reverse direction, combined effect of these two results in reduction of the net dispersion coefficient. However, in presence of an external thermal gradient, the induced thermo-electric streaming field may aid or oppose the pressure gradient induced streaming field depending on different fluidic conditions or parameters as discussed earlier. A typical variation of dispersion coefficient ratio ($\bar{D}_{eff}$) with $C_\mu$ is illustrated in figure 5d (i). Evaluated at $\gamma$ = 0.1, $\bar{D}_{eff}$ obeys linear relationship with $C_\mu$ at the initial stages (within $1 \le C_\mu \le 2$), then increases abruptly to experience a pronounced amplification of dispersion coefficient ratio as an increment of ~ 4.5 times in $\bar{D}_{eff}$ with $C_\mu$ can be seen from figure 5d (i).

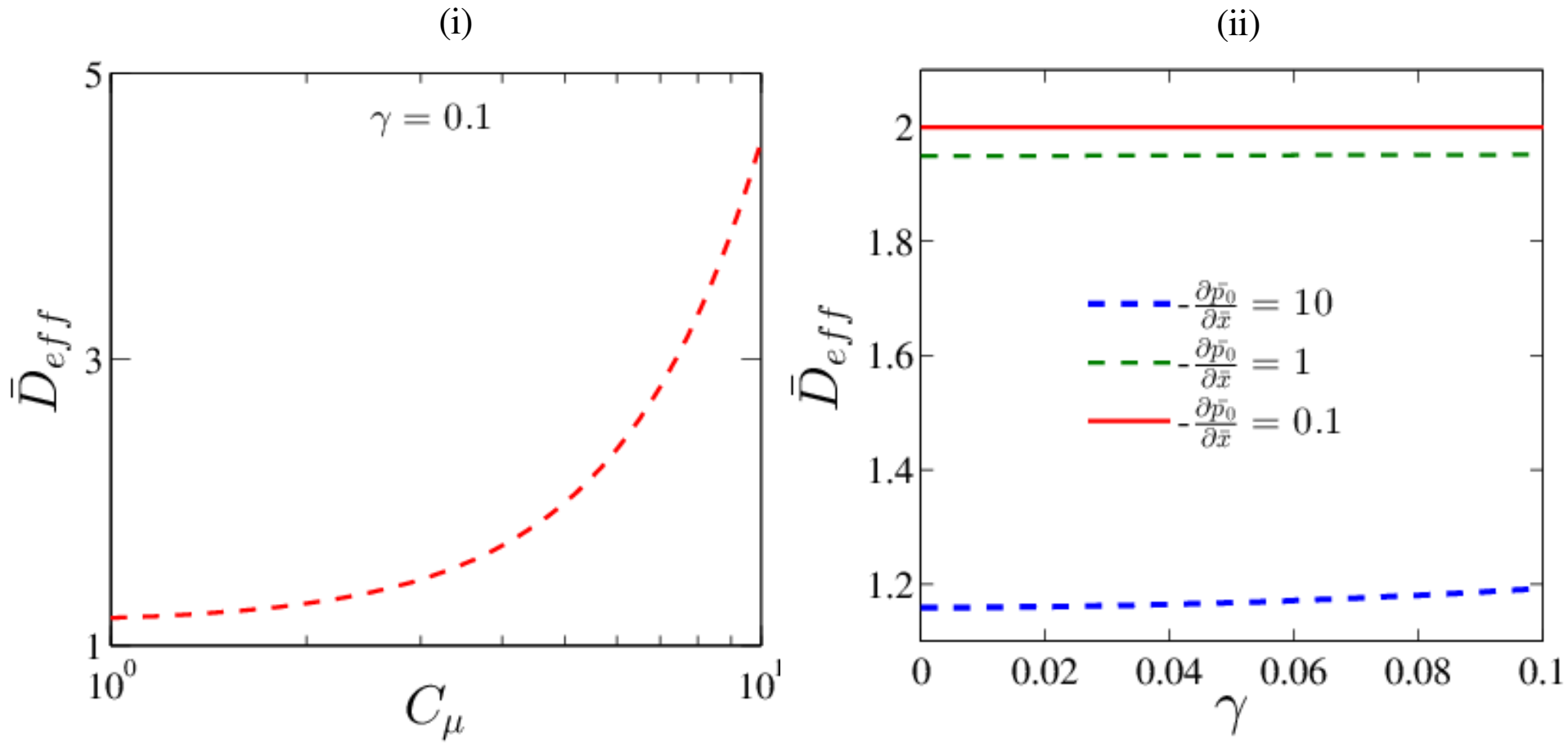

FIGURE 5d. (i) Dependence of the dispersion coefficient ratio ($\bar{D}_{eff}$) with $C_\mu$, (ii) variation of the same with $\gamma$ for varying strength of imposed pressure gradient.

The variation of $\bar{D}_{eff}$ with $\gamma$ is shown in figure 5d (ii) where $\gamma$ is the ratio of imposed temperature difference to the cold side temperature which determines the degree of thermal perturbation to the flow field. In purely $\Delta T$ driven flow, increasing $\gamma$ significantly alters the thermo-electric field and the resulting flow dynamics and accordingly $\bar{D}_{eff}$ should show strong dependence with $\gamma$. Introducing pressure-gradient restricts this sensitivity of $\bar{D}_{eff}$ with $\gamma$ where $\bar{D}_{eff}$ increases slightly from ~ 1.16 times to ~ 1.19 times as $\gamma$ is changing from 0 to 0.1. Increasing pressure gradient results in enhancing the magnitude of the flow velocity and dispersion coefficient with the influence of $\gamma$ becomes increasingly insignificant.

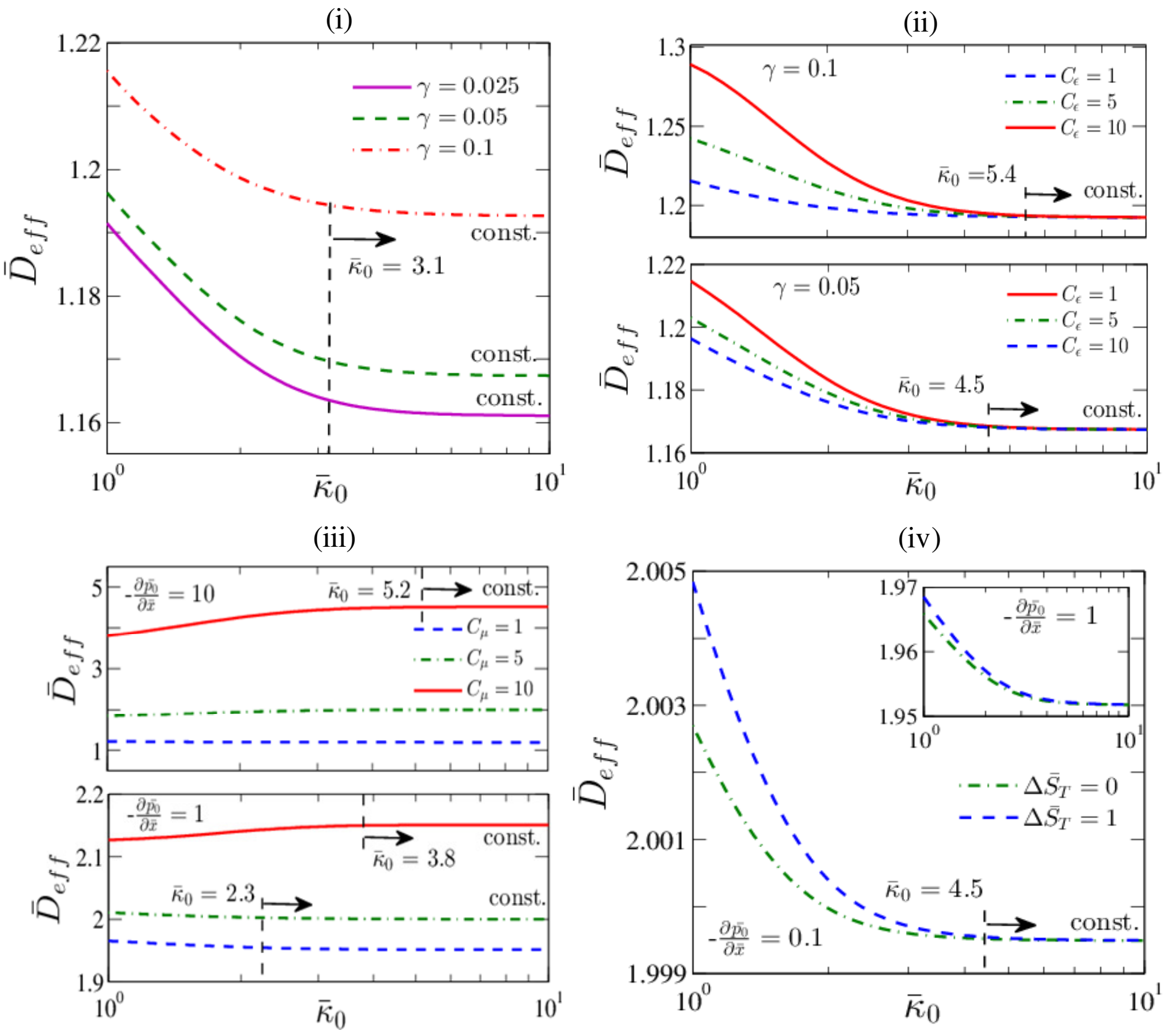

FIGURE 5e. Dispersion coefficient ratio ($\bar{D}_{eff}$) with $\bar{\kappa}_0$ for different (i) $\gamma$, (ii) $C_\varepsilon$, (iii) $C_\mu$ and (iv) $\Delta\bar{S}_T$ respectively.

Another important parameter in influencing the flow physics is the degree of channel confinement $(\bar{\kappa}_0)$, depending on which the interacting forces may become predominant or become insignificant. The effect of $\bar{\kappa}_0$ on dispersion coefficient ratio ($\bar{D}_{eff}$) has been demonstrated in figure 5e. Irrespective of the degree of thermal perturbation (value of $\gamma$), here $\bar{D}_{eff}$ decreases gradually with increasing $\bar{\kappa}_0$ and beyond a critical value of $\bar{\kappa}_0$, it reaches a constant value where this magnitude becomes higher with increasing $\gamma$ with saturation occurring relatively earlier (figure 5e (i)). The reduction in the streaming current and resulting streaming potential with increasing $C_\varepsilon$ leads to an increment in dispersion coefficient as $\bar{D}_{eff}$ rises to ~ 1.29 from 1.22 as $C_\varepsilon$ is changed from1 to 10 (evaluated at $\gamma$ = 0.1, as shown in figure 5e (ii)). The dependence with $\bar{\kappa}_0$ for varying $C_\varepsilon$ is similar to figure 5e (i) where after decaying gradually, $\bar{D}_{eff}$ approaches to a constant value at higher $\bar{\kappa}_0$ with saturation occurring later at higher $\gamma$. Also, at higher $\gamma$, more is the influence of $C_\varepsilon$ on dispersion coefficient because of strengthened thermo-electric perturbation. In figure 5e (iii), variation of the same with $\bar{\kappa}_0$ is shown with varying $C_\mu$ for two different strengths of pressure gradient. For lower strength $(-\partial\bar{p}_0/\partial\bar{x}=1)$, $\bar{D}_{eff}$ first decreases with $\bar{\kappa}_0$ for lower $C_\mu$ and approaches a constant value beyond $\bar{\kappa}_0$ = 2.3 while role reversal has been observed for higher $C_\mu$ (with saturation occurring later) because of pronounced reduction of flow resistance. For higher strength $(-\partial\bar{p}_0/\partial\bar{x}=10)$, the variation of $\bar{D}_{eff}$ with $\bar{\kappa}_0$ remains unaffected for lower $C_\mu$ while at higher $C_\mu$, after increasing gradually with $\bar{\kappa}_0$, $\bar{D}_{eff}$ reaches a constant value of ~ 4.5 later at $\bar{\kappa}_0$ = 5.2. Overall, the magnitude of $\bar{D}_{eff}$ always remains much higher compared to unity because of easier actuation of flow owing to lesser viscous resistance. At lower strength of pressure gradient (i.e. $-\partial\bar{p}_0/\partial\bar{x}=1$), the reduction of streaming potential with thermophoretic mobility difference ($\Delta\bar{S}_T$) is reflected through the enhancement of $\bar{D}_{eff}$ with increasing $\Delta\bar{S}_T$ as depicted in figure 5e (iv) although the influence is very weak. At higher strength (shown in the inset of figure 5e (iv)), trends are similar where the magnitude of $\bar{D}_{eff}$ gets reduced from ~ 2.005 to ~ 1.97 as visible in figure 5e (iv).

### 3.2 *Effect of transverse temperature gradient*

Here the primary driving force for flow actuation is the axially applied pressure gradient while the contribution from the imposed thermal gradient ($\Delta T$) is secondary through physical property alteration and introducing a permittivity-variation induced body force (this term is $C_{\varepsilon}\dfrac{\partial\overline{\varepsilon}}{\partial\overline{y}}\dfrac{\partial\theta}{\partial\overline{y}}$ as observed in the $x$-component of momentum equation). Apart from this, a concentration gradient in the transverse direction is also induced because of the imposed temperature difference which creates a transverse migration of the ions from the hot region to the cold region (mathematically this can be observed from equation (33) which tells us that in the electro-neutral (i.e. $\phi$ = 0) region, $\dfrac{1}{\overline{n}_i}\dfrac{\partial\overline{n}_i}{\partial\overline{y}} = -\overline{S}_{Ti}\,\gamma\dfrac{\partial\theta}{\partial\overline{y}}$). This migration acts as a resistance in the axial separation between the ions thus creating weaker streaming field. Further source of alteration in the streaming potential upon the applied thermal gradient comes from factors like modulated viscous resistance of flow, change in effective EDL thickness, permittivity variation induced alteration in streaming current (as previously discussed) etc.

Contrary to the case of axial thermal gradient, here the flow physics is strongly dependent on the parameter $\chi$ which indicates the difference in diffusivities between the counter-ions and co-ions. Decreasing the value of $\chi$ indicates more diffusivity of the co-ions than counter-ions in the upstream section. This leads to more migration of the co-ions in the upstream section, resulting in a reduction of the streaming potential. As depicted in inset I of figure 6a (i), streaming potential ratio ($Er$) decreases twice from 4 to 2 as $\chi$ is changing from -0.1 to -0.3. As $\chi$ decreases, the magnitude of the maximum velocity increases more than twice, as observed in figure 6a (i). Interestingly, for $\chi$ = -0.3, the position of maxima is close to the channel centreline while for $\chi$ = -0.1, this is shifted away from the channel centreline towards the right i..e. towards the direction of thermal gradient. For $\chi$ = -0.1, velocity profile is slightly deviating from the parabolic behavior because of the generation of stronger streaming field induced back flow. As $\chi$ is decreasing, lesser generation of streaming potential makes the effect of pressure gradient predominant and velocity distribution follows parabolic behavior. Here, the effect of $\chi$ on the flow field is observed only in the higher degree of confinement of channel (observed at $\overline{\kappa}_0 = 1$) and gets diminished with decreasing the extent of confinement (i.e. larger region of electro-neutrality). The inset II of figure 6a (i) shows that increasing $\overline{\kappa}_0$ (i.e. decreasing confinement) from 1 to 10 makes the effect of $\chi$ on flow-field insignificant. With decreasing $\chi$, there is a negligible difference between the magnitude of maximum velocity as shown by zoomed view in inset II.

Keeping $\chi$ constant, the velocity distribution for varying $C_\mu$ is shown in figure 6a (ii). Since $C_\mu$ signifies the sensitivity of fluid viscosity with temperature, higher the value of $C_\mu$, higher is the reduction of the viscosity, lower is the resistance to drive the flow. Because of the strengthened advective current, streaming potential ratio ($E_r$) is increased slightly as $C_\mu$ is increased from 1 to 5. For lower $C_\mu$, the velocity profile is parabolic in nature while at higher $C_\mu$, the enhanced sensitivity of fluid viscosity creates strong departure from the parabolic distribution with the maxima being shifted towards the hot region. Increasing $\bar{\kappa}_0$ ensures lesser penetration of EDL in the bulk resulting in lowering the net streaming potential thereby enhancing the magnitude of the flow velocity to some extent as depicted by the dotted lines in figure 6a (ii).

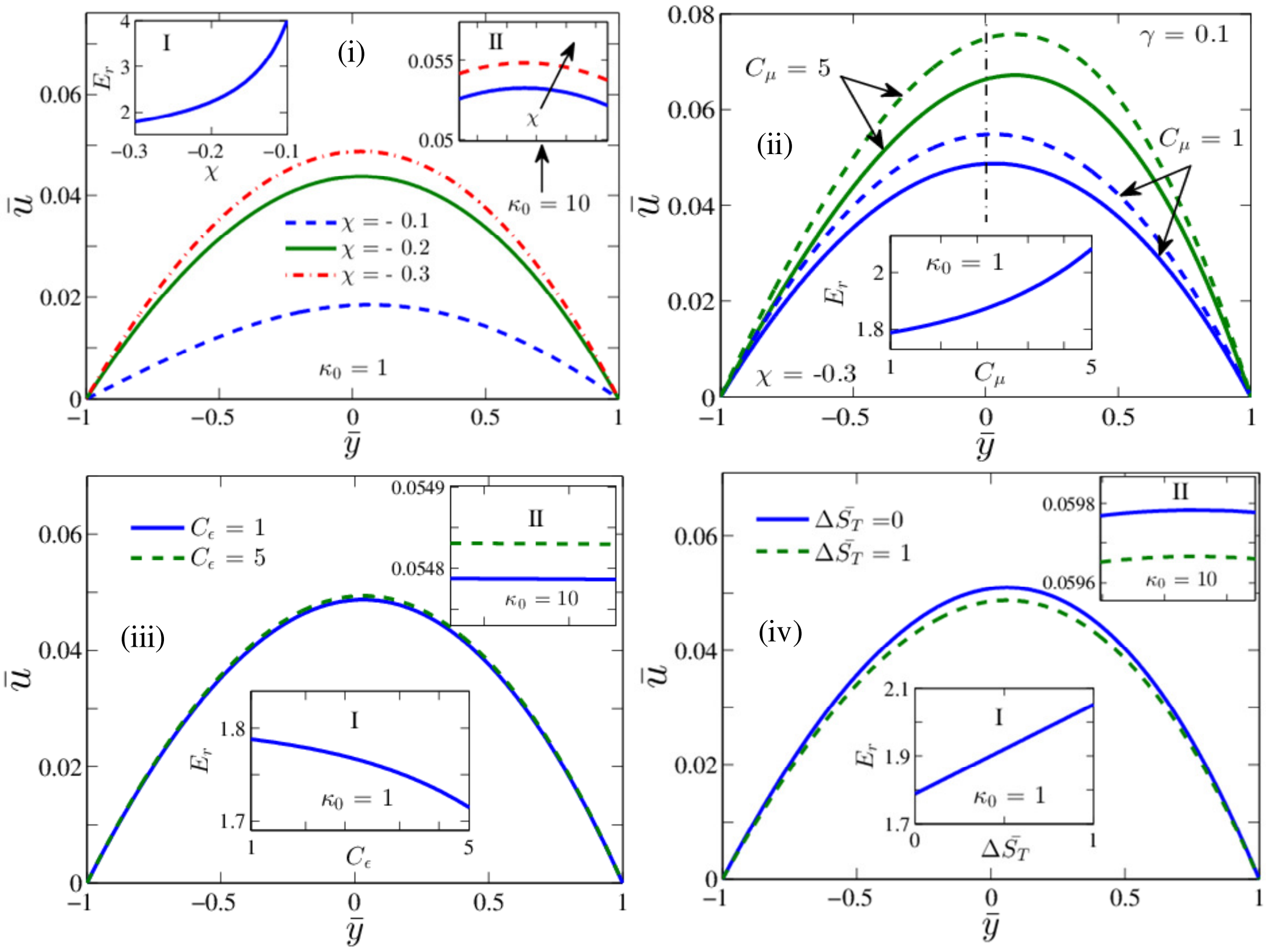

FIGURE 6a. Velocity distribution in the $y$-direction for varying (i) $\chi$, (ii) $C_\mu$, (iii) $C_\varepsilon$ and (iv) $\Delta\bar{S}_T$ respectively (evaluated at $-\partial\bar{p}_0/\partial\bar{x} = 0.1$).

Figure 6a (iii) shows the transverse variation of the velocity field with increasing $C_\varepsilon$. Here, $C_\varepsilon$ indicates the sensitivity of permittivity with temperature. The contribution of $C_\varepsilon$ primarily comes through the alteration in the potential distribution upon increasing $C_\varepsilon$ and through the permittivity-induced forcing term in the fluid momentum transport. Here, streaming potential ratio ($E_r$) decreases (inset I of figure 6a (iii)) slightly with increasing $C_\varepsilon$ resulting small increment in the velocity magnitude. Similar to the effect of $\chi$, here also increasing $\bar{\kappa}_0$ from 1 to 10 makes the effect of $C_\varepsilon$ on the flow field inconsequential as evident from inset II (where zoomed view of maximum velocity is presented) of figure 6a (iii).

The variation of the velocity field in the $y$-direction for different $\Delta\bar{S}_T$ is shown in figure 6a (iv) where $\Delta\bar{S}_T$ means the difference in thermophoretic mobilities between cations and anions. Increasing $\Delta\bar{S}_T$ implies higher thermophoretic mobility of counter-ions than co-ions which leads to preferential migration of the counter-ions towards the cold region. This leads to an ionic redistribution resulting an asymmetry in the potential distribution within EDL. On observing the momentum equation, one can understand that the role of $\Delta\bar{S}_T$ comes through the charge distribution alteration (via the modulated EDL thickness and the term $\left\{C_\varepsilon - \left(1 + \bar{S}_{Tavg}\right)\right\} \bar{\kappa}_0^2\, \theta_0\, \bar{\psi}_0$) which further perturbs the fluid advection motion. This in turn influences the advection current and the induced streaming field. Accordingly, the streaming potential ratio ($E_r$) increases from ~ 1.8 times to ~ 2.05 times as one increases $\Delta\bar{S}_T$ from 0 to 1 following a linear dependence. The enhanced streaming field induces more backward flow thus lowering the magnitude of the flow velocity as observed in figure 6a (iv) for $\Delta\bar{S}_T = 1$.

As previously discussed in the case of axial thermal gradient, any perturbation to the flow field is strongly reflected in the associated dispersion characteristics because of its strong depedence on the non-uniformity of the flow velocity. By inspecting the velocity distributions demonstrated by figure 6a, one general observation can be made that the effect of most of the parameters (except $C_\mu$) on flow field is noticeable only under higher confinement (i.e. at lower $\bar{\kappa}_0$) and becomes negligible at higher $\bar{\kappa}_0$. Apart from $\bar{\kappa}_0$, $C_\mu$ is another important parameter whose effect on flow field is far more significant compared to other parameters ($C_\varepsilon$, $\Delta\bar{S}_T$) not only by altering the magnitude of the flow velocity but also creating strong departure from the parabolic distribution. Accordingly, in figure 6b we have incorporated the variation of dispersion coefficient ratio ($\bar{D}_{eff}$) with two parameters $C_\mu$ and $\bar{\kappa}_0$. Similar to axial $\Delta T$, drastic reduction in flow resistance with increasing $C_\mu$ is reflected in figure 6b (i) where $\bar{D}_{eff}$ is increased up to 3 times as $C_\mu$ is varying from 1 to 10.

The reduced strength of electrokinetic forces with increasing $\bar{\kappa}_0$ lowers the thermo-electric perturbation to the flow field and hence, dispersion coefficient ratio ($\bar{D}_{eff}$) (at $\gamma = 0.1$) reduces from 1.19 to 1.18 as $\bar{\kappa}_0$ is changed from 1 to 10 with alteration being suppressed beyond $\bar{\kappa}_0 = 4.4$. Corresponding results for lower $\gamma$ are similar with saturation occuring at higher $\bar{\kappa}_0$.

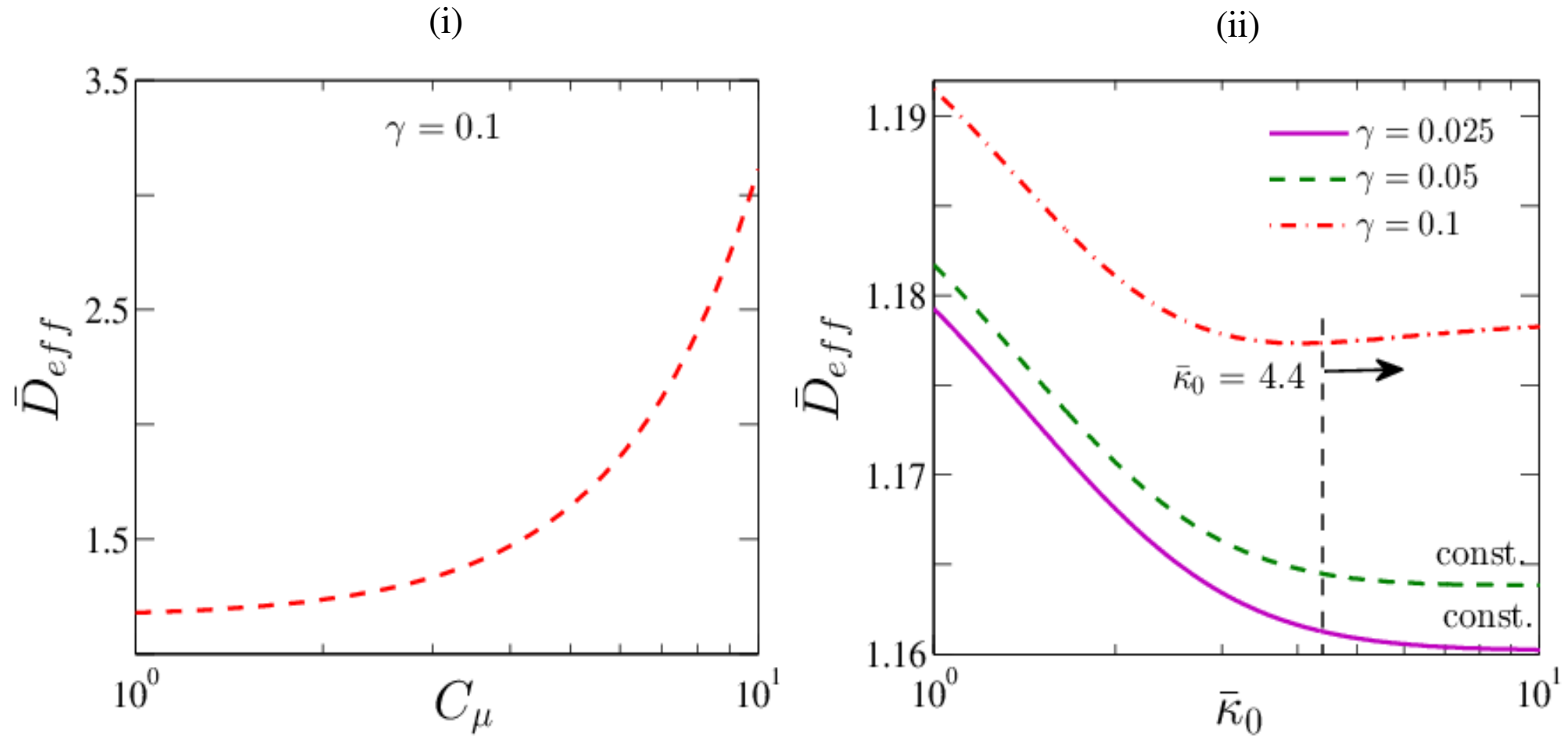

FIGURE 6b. (i) Variation of dispersion coefficient ratio ($\bar{D}_{eff}$) with $C_\mu$, (ii) variation of the same with $\bar{\kappa}_0$ for different $\gamma$.

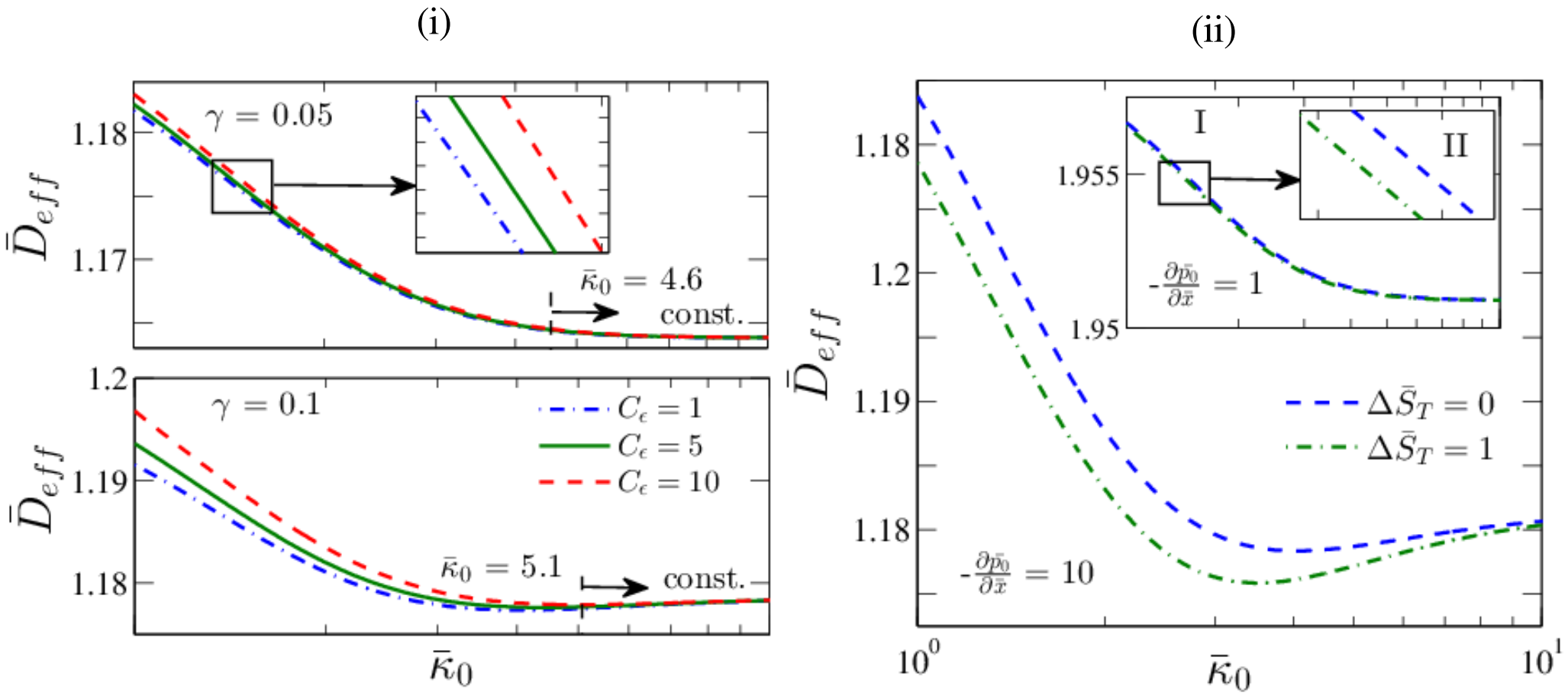

FIGURE 6c. Variation of $\bar{D}_{eff}$ with $\bar{\kappa}_0$ for different values of (i) $C_\varepsilon$ and (ii) $\Delta\bar{S}_T$ respectively.

Now, the effect of two other parameters $C_\varepsilon$ and $\Delta\overline{S}_T$ are shown in figure 6c where the reduced streaming potential with increasing $C_\varepsilon$ results in increasing the dispersion coefficient. However, $C_\varepsilon$ turns out to be less effective at lower thermal perturbation ($\gamma = 0.05$) and noticeable effect can only be observed at higher $\gamma$ (figure 6c (i)). Similarly, higher volumetric suppression due to enhanced streaming potential with increasing $\Delta\overline{S}_T$ lowers the dispersion coeffficient with $\Delta\overline{S}_T$ becoming influential only at higher strength of pressure gradient (figure 6c (ii)).

### 3.3 *Effect of fluid rheology*

Now, the inclusion of rheological aspect of fluid on streaming potential is highlighted in figure 7a where the variation of the streaming potential ratio ($E_r$) is shown with Deborah number ($De$) in case of an axially applied thermal gradient. For simplicity of analysis, here we have chosen dilute polymeric solution mixed with electrolyte (aqueous solutions of well-known polymers like

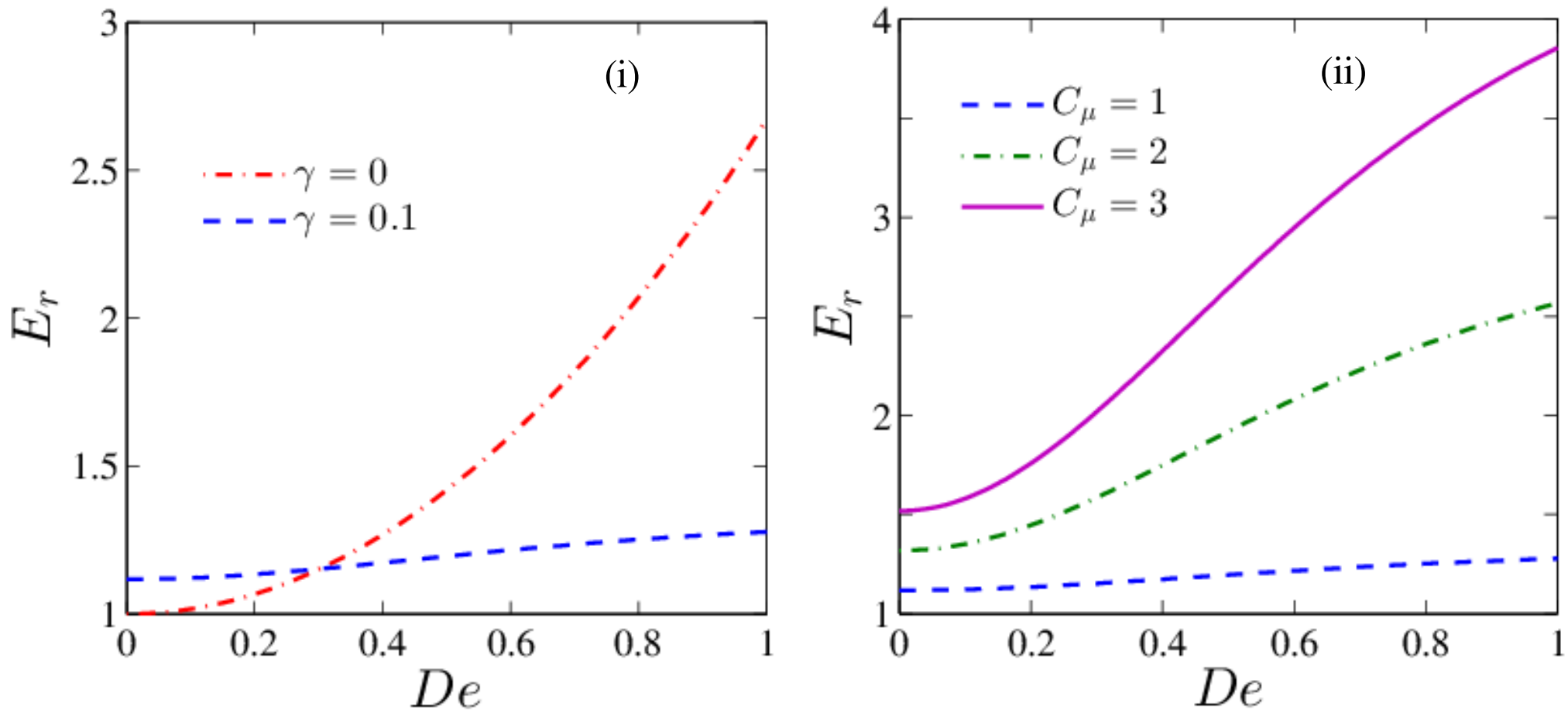

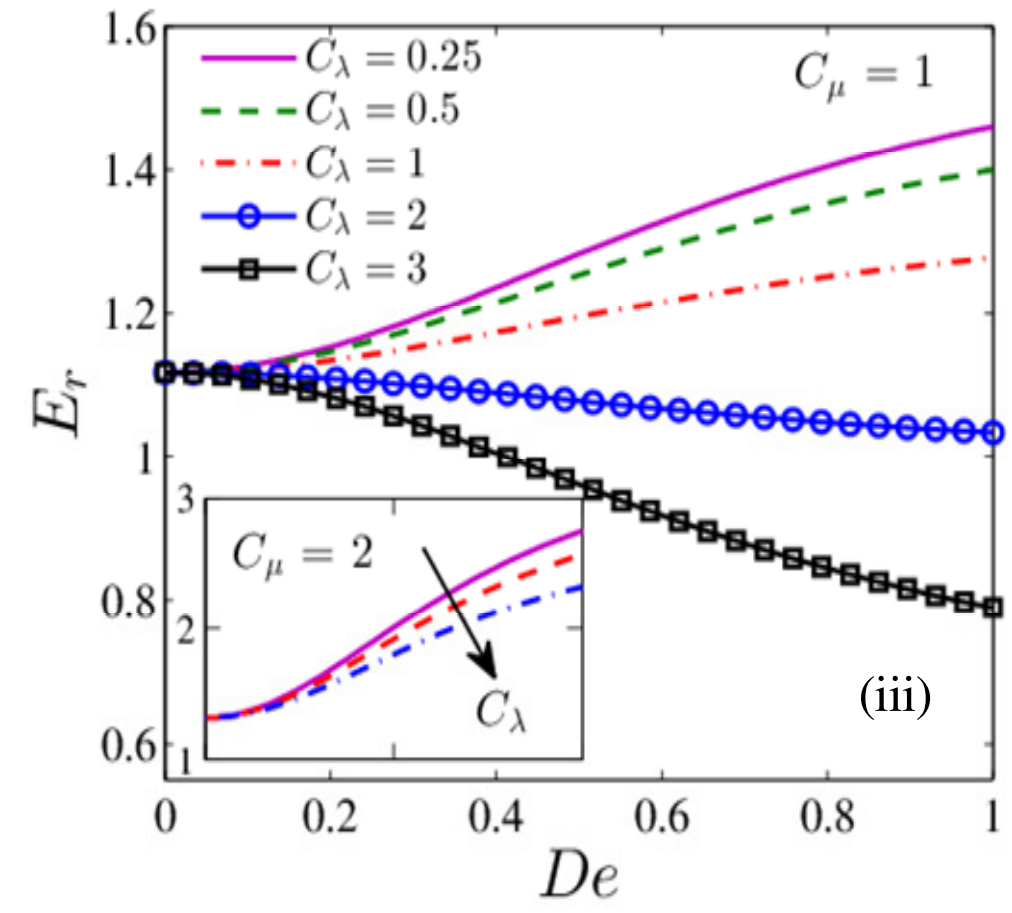

FIGURE 7a. Streaming potential ratio ($E_r$) as a function of Deborah number ($De$) for different (i) $\gamma$, (ii) $C_\mu$ and (iii) $C_\lambda$ respectively.

Polyethylene oxide (PEO), Polyacrylamide (PAM) can be taken as example) as a reference viscoelastic fluid. If the polymer concentration remains below a certain threshold concentration (commonly known as cross-over concentration or overlap concentration), no interaction between the polymer chains can take place and we can assume the solution to belong within the dilute regime (Del Giudice *et al.* 2015, 2017; Tirtaatmadja *et al.* 2006). For dilute solutions, the imposed driving force creates a disturbance on the polymer chain, in response of which there is an expansion of the polymer chain (Larson 2005). This expansion in turn returns a disturbance to the flow field thus influencing the velocity distribution. Here, it is necessary to highlight the significance of Deborah number ($De$) which determines the relative strength between elastic and viscous effects. Higher the value of $De$, higher is the extent of viscoelasticity which can be attributed to either increased elasticity of fluid (thus creating more disturbance in the flow field) or attenuated viscous resistance in the flow (because of pronounced shear-thinning effect). For purely pressure-driven flow ($\gamma = 0$, i.e. isothermal condition), elastic behavior of fluid remains unaffected and increasing $De$ causes significantly amplified shear-thinning effect which facilitates the fluid advective motion and therefore, streaming potential ratio ($E_r$) increases up to ~ 2.7 times as compared to a Newtonian fluid (i.e. $De = 0$). Now, as thermal gradient is imposed, degree of viscoelasticity gets strongly influenced as fluid viscosity and relaxation time both becomes strong function of temperature. Here, it is noteworthy to mention that in the dilute regime, the relaxation time of a polymeric solution remains independent of the polymer concentration, described by widely known Zimm's relaxation time ($\lambda_z$) (Del Giudice *et al.* 2017; Pan *et al.* 2018; Tirtaatmadja *et al.* 2006). Previous experimental studies have reported an inverse relationship of $\lambda_z$ with temperature which can be approximated by an exponential thinning behavior (i.e. in the form of $\lambda_{eff} = \lambda_{ref} \exp\left[-\omega_4\left(T - T_{ref}\right)\right]$). So, the net effect of viscoelasticity on streaming potential depends on the relative sensitivity of viscosity ($C_\mu$) and relaxation time ($C_\lambda$) of fluid with temperature. For a fixed value of $C_\mu$ and $C_\lambda$, introducing thermal gradient ($\gamma = 0.1$) results a reduction in the net streaming potential with $De$ where an increment of ~ 1.27 times in $E_r$ is observed as opposed to ~ 2.7 times increment for $\gamma = 0$ (figure 7a (i)). Interestingly, a cross-over at $De = 0.3$ takes place between the graphs of $\gamma = 0$ and $\gamma = 0.1$. Below this critical $De$, the magnitude of the streaming potential is higher for combined temperature gradient and pressure-driven flow and beyond $De = 0.3$, it falls below the streaming potential in isothermal condition in which case the strongly pronounced shear thinning effect with increasing $De$ dictates the flow physics creating faster rise in streaming potential. Now, at lower $C_\mu$ (denoting lower temperature-sensitivity of fluid viscosity), this two aforesaid counteracting factors are of comparable magnitude resulting in a slight increase in $E_r$ (from 1.12 times to 1.28 times) as *De* is varying from 0 to 1, visible in figure 7a (ii). With increasing $C_\mu$, its effect starts to become dominant over $C_\lambda$ and $E_r$ undergoes significant enhancement up to ~ 3.86 times (at $C_\mu = 3$) compared to the Newtonian fluid. Similarly, increasing $C_\lambda$ from 0.25 leads to faster reduction in fluid relaxation time denoting elevated elasticity-mediated disturbance to the axial separation between ions thus lowering the net streaming potential. As evident from figure 7a (iii), $E_r$ decreases from 1.46 to 0.79 as $C_\lambda$ is increased from 0.25 to 3. Beyond $C_\lambda = 1$, $E_r$ starts to fall with *De* (from its reference value ~ 1.12) and becomes less than unity at higher *De*. This tells us that if $C_\lambda$ is high, (i.e. rapid reduction of fluid relaxation time with temperature) employing a Newtonian fluid is more advantageous instead of a viscoelastic fluid as far as streaming potential generation is concerned. Now, as shown in the inset of figure 7a (ii), the influence of $C_\lambda$ on streaming potential gets dampened at higher $C_\mu$ ($C_\mu = 2$) where $E_r$ reduces at a lower rate from 2.75 times to 2.3 times (with respect to a Newtonian fluid) with increasing $C_\lambda$.

The flow rate ratio ($Q_r$) variation with Deborah number (*De*) is highlighted in figure 7b for two varying factors $C_\mu$ and $C_\lambda$ respectively. At one side, increasing $C_\mu$ induces more streaming potential leading to net volumetric suppression due to reverse electrokinetic flow, on the other hand, viscous resistance in the flow decreases significantly. Hence, net throughput through the microchannel is determined by their relative strengths. For lower value of $C_\mu$, $Q_r$ remains almost unaffected with the variation of *De* (at $C_\mu = 1$). However, with increasing $C_\mu$, $Q_r$ increases sharply with an enhancement up to ~ 2.03 times can be noticed at $C_\mu = 3$ (figure 7b (i)).

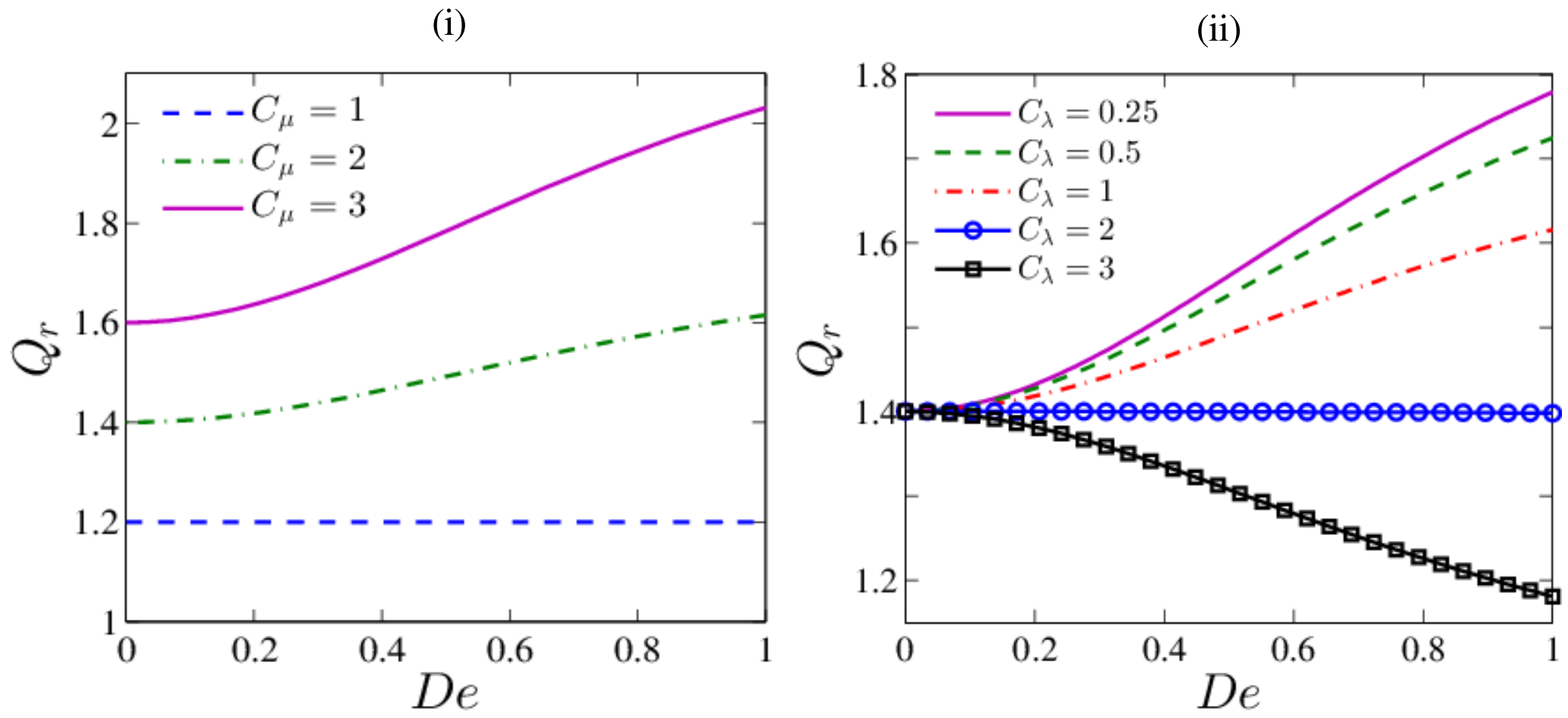

FIGURE 7b. Volume flow rate ratio ($Q_r$) with *De* for different (i) $C_\mu$ and (ii) $C_\lambda$ respectively.

Similarly, increasing $C_\lambda$ has inverse effect on the flow rate as $Q_r$ decreases with $De$ from 1.78 times to 1.18 times as $C_\lambda$ is changed from 0.25 to 3. Since the net streaming potential reduces beyond $C_\lambda = 1$, reduced volumetric suppression makes $Q_r$ still higher than unity at $C_\lambda = 3$.

This section has reached its culmination where the dependence of the dispersion coefficient ratio ($\bar{D}_{eff}$) is shown with Deborah number ($De$) for two crucial parameters $C_\mu$ and $C_\lambda$. Similar to the variation of $Q_r$, $\bar{D}_{eff}$ is also highly sensitive to the variation in fluid viscosity and relaxation time. Looking into the constitutive form of a viscoelastic fluid one can realize that the inherent non-linearity in stress-tensor terms gets amplified with increasing $C_\mu$. Physically, the impact of physical property alteration is reflected more in viscoelastic fluids compared to Newtonian fluids and accordingly, $\bar{D}_{eff}$ increases up to 1.9 times at higher $C_\mu$ (at $C_\mu = 3$). Also, changing $C_\lambda$ can result significant alteration in the dispersion coefficient where $\bar{D}_{eff}$ exhibits an inverse dependence on Deborah number ($De$) at higher $C_\lambda$, as observed in figure 7c (ii). Overall, the dispersion coefficient ratio ($\bar{D}_{eff}$) undergoes a reduction from 1.53 to 1.1 as $C_\lambda$ is increased from 0.25 to 3.

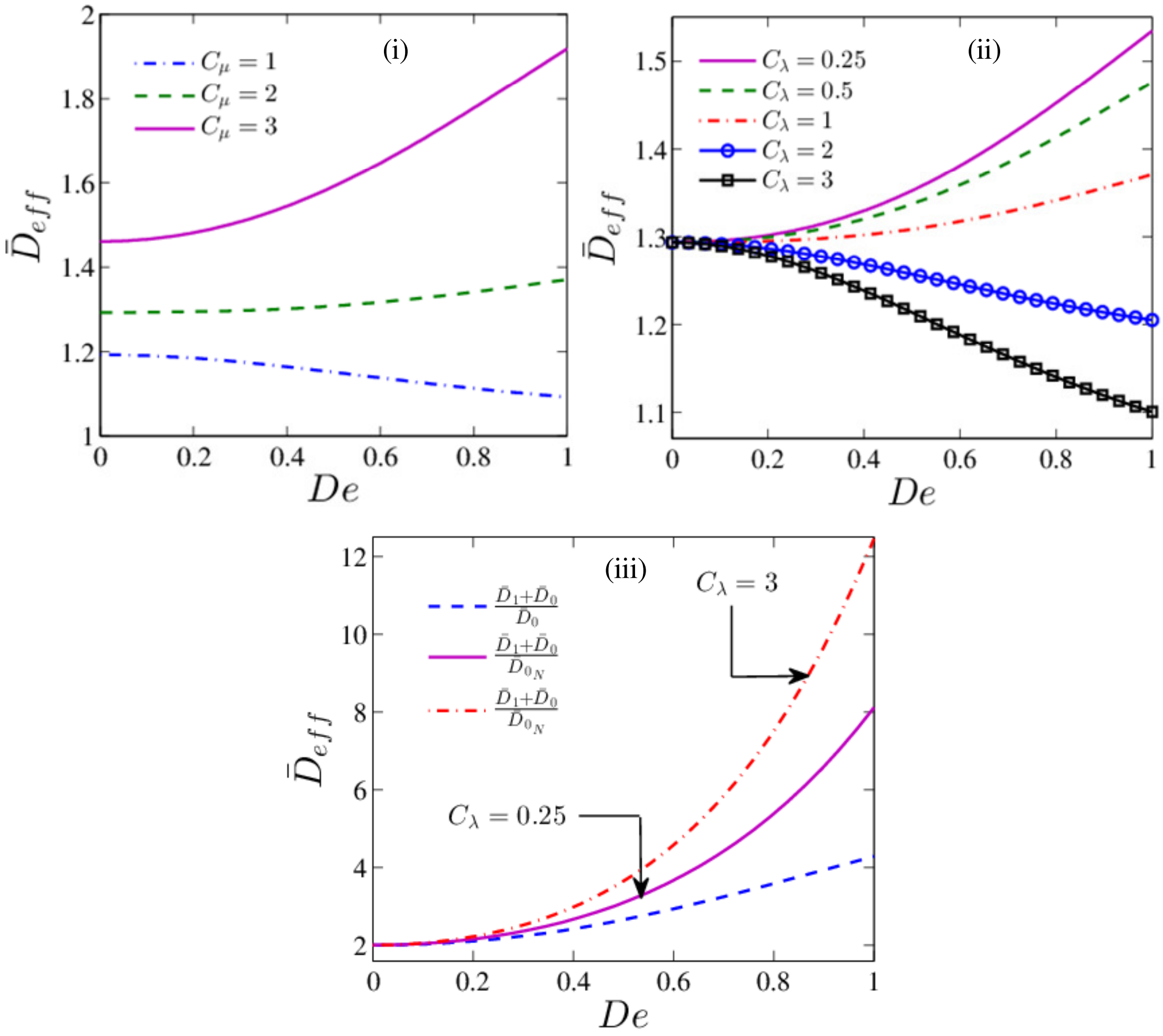

FIGURE 7c. Variation of dispersion coefficient ratio ($\bar{D}_{eff}$) with Deborah number ($De$) for (i) different $C_\mu$ and (ii) different $C_\lambda$. (iii) Variation of $\bar{D}_{eff}$ for specific combination of $C_\mu$ and $C_\lambda$.

From the previous figure, it is clearly evident that increasing $C_\mu$ and decreasing $C_\lambda$ turns out to be favorable as far as the enhancement of dispersion is concerned. When we combine this two factors, under strong thermal perturbation (at $\gamma$ = 0.1), the variation of the dispersion coefficient ratio ($\bar{D}_{eff}$) with Deborah number ($De$) experiences massive augmentation. The dotted blue line represents the ratio of relative increment of $\bar{D}_{eff}$ in viscoelastic fluid for combined pressure gradient and thermal gradient driven flow as compared to that for a purely pressure-driven flow where $\bar{D}_{eff}$ is increased up to ~ 4.3 times. Now, the dash-dot line (red colored) expresses the ratio of the net dispersion coefficient in viscoelastic fluid to that compared to the solely pressure-driven flow of a Newtonian fluid where it augments further up to ~12.5 times as $De$ is varied from 0 to 1. Keeping in mind its applicability under actual experimental conditions, we now focus on the physically relevant values of $C_\mu$ and $C_\lambda$. As reported in the literature, the relative change of fluid viscosity with temperature $(1/\mu)(\partial\mu/\partial T)$ for electrolyte solutions is approximately ~ $15\times10^{-3}$ $K^{-1}$ (Dietzel & Hardt 2017) which can be used in the expression of $\bar{\mu} = \mu/\mu_{ref} = \exp(-\gamma C_\mu \theta)$ where the value of $C_\mu$ turns out to be close to 5. Also the reduction of fluid relaxation time with temperature can be correlated in the similar fashion ($\bar{\lambda} = \lambda_{eff}/\lambda_{ref} = \exp(-\gamma C_\lambda \theta)$) where the value of $C_\lambda$ is chosen as 3 (the reason behind this particular combination of the parameters is discussed in Section E1 of the supplementary material). For $C_\mu$ = 5, $C_\lambda$ = 3, $\bar{D}_{eff}$ is increased up to ~ 8.1 times as compared to a Newtonian fluid. Therefore, we conclude that, employing this combination of the parameters under the combinatorial effect of external pressure gradient and temperature difference, it is indeed practically possible to achieve up to one order of magnitude (approximately) enhancement in the dispersion coefficient. For completeness of the present analysis, the results for transverse thermal gradient are presented in Section E2 of the supplementary material.

## 4. Conclusions

We have considered thermally-modulated elctrokinetic transport to realize significant enhancement in solute dispersion of a complex fluid through a microfluidic channel. Although several techniques in the past have been deployed towards modulating the uniform velocity profile of electrically actuated flows, improved hydrodynamic dispersion still remains unexplored. In this context, the present study shows that combining the interplay between thermal and electrical effects coupled with fluid rheology, one can achieve up to one order of magnitude enhancement of dispersion coefficient in a pressure-driven flow of an electrolyte. This is mediated by breaking the equilibrium of the charge distribution within the electrical double layer upon imposed thermal gradient, subsequent modulation in thermo-physical properties, and eventual alterations in the fluid motion. We believe that such complex coupling between thermal, electrical, hydro-dynamic and rheological parameters in small scales can be exploited to a benefit in the design of thermally-actuated micro and bio-microfluidic devices demanding improved hydrodynamic dispersion.

**Acknowledgement**

SC acknowledges Department of Science and Technology, Government of India, for Sir J. C. Bose National Fellowship.